\pdfoutput=1
\documentclass[
    aps,pre,twocolumn,
    reprint,
    superscriptaddress,
    nofootinbib,
    floatfix,
    amssymb,
    longbibliography
]{revtex4-2}

\usepackage{amsmath}
\usepackage{amsfonts}
\usepackage{amssymb}
\usepackage{graphicx}
\usepackage{bm}

\usepackage[utf8]{inputenc}

\usepackage{nicefrac}

\usepackage[normalem]{ulem}

\usepackage{xcolor}

\definecolor{linkColor}{rgb}{0,0.3,0.7}
\usepackage{hyperref}
\hypersetup{colorlinks=true,
            allcolors=linkColor,
            pdfborder={0 0 0},
            pdfencoding = auto
            }

\def\etaStat{\eta_\mathrm{stat}}
\def\rhoStat{\rho_\mathrm{stat}}
\def\cStat{c_\mathrm{stat}}

\begin{document}

\title{Coarsening dynamics of chemotactic aggregates}

\author{Henrik Weyer}
\thanks{H.W.\ and D.M.\ contributed equally to this work.}
\author{David Muramatsu}
\thanks{H.W.\ and D.M.\ contributed equally to this work.}
\affiliation{Arnold Sommerfeld Center for Theoretical Physics and Center for NanoScience, Department of Physics, Ludwig-Maximilians-Universit\"at M\"unchen, Theresienstra\ss e 37, D-80333 M\"unchen, Germany}
\author{Erwin Frey}
\email{frey@lmu.de}
\affiliation{Arnold Sommerfeld Center for Theoretical Physics and Center for NanoScience, Department of Physics, Ludwig-Maximilians-Universit\"at M\"unchen, Theresienstra\ss e 37, D-80333 M\"unchen, Germany}
\affiliation{Max Planck School Matter to Life, Hofgartenstraße 8, D-80539 Munich, Germany}

\date{May 23, 2025}
	
\begin{abstract}
Auto-chemotaxis, the directed movement of cells along gradients in chemicals they secrete, is central to the formation of complex spatiotemporal patterns in biological systems.
Since the introduction of the Keller--Segel model, numerous variants have been analyzed, revealing phenomena such as coarsening of aggregates, stable aggregate sizes, and spatiotemporally chaotic dynamics.
Here, we consider general mass-conserving Keller--Segel models, that is, models without cell growth and death, and analyze the generic long-time dynamics of the chemotactic aggregates.
Building on and extending our previous work, which demonstrated that chemotactic aggregation can be understood through a generalized Maxwell construction balancing density fluxes and reactive turnover, we use singular perturbation theory to derive the rates of mass competition between well-separated aggregates.
We analyze how this mass-competition process drives coarsening in both diffusion- and reaction-limited regimes, with the diffusion-limited rate aligning with our previous quasi-steady-state analyses.
Our results generalize earlier mathematical findings, demonstrating that coarsening is driven by self-amplifying mass transport and aggregate coalescence.
Additionally, we provide a linear stability analysis of the lateral instability, predicting it through a nullcline-slope criterion that parallels the curvature criterion in spinodal decomposition.
Overall, our findings suggest that chemotactic aggregates behave similarly to phase-separating droplets, providing a robust framework for understanding the coarse-grained dynamics of auto-chemotactic cell populations and a quantitative basis for comparing chemotactic coarsening to canonical non-equilibrium phase separation.
\end{abstract}

\maketitle

\section{Introduction}
\label{sec:introduction}

Chemotaxis is a fundamental biological strategy that enables the directed and collective movement of individual cells. This process involves cells detecting signaling molecules and navigating along their concentration gradients. 
Chemotaxis plays a crucial role in various biological systems, including bacterial colonies \cite{Berg2004}, tissue organization \cite{Murray2003,Roussos.etal2011,Trepat.etal2012}, or the immune system \cite{Gerard.Rollins2001}.
Chemotactic-like processes extend beyond natural biological systems, playing a pivotal role in synthetic 
\cite{Ebbens.Howse2010}
and reconstituted systems \cite{Ramm.etal2021}. 
For example, different types of phoretic processes drive the directed movement of colloidal particles in response to (self-generated) solute concentration gradients \cite{Ebbens.Howse2010,Liebchen.Lowen2018,Stark2018,Golestanian2022}.
Moreover, cross-diffusion of enzymes with their substrates and products, allows that catalysis-driven fluxes induced by the enzymes result in their aggregation \cite{Cotton.etal2022}.
Finally, enzyme-enriched condensates interacting with the substrates and products experience chemotaxis-like cross-diffusion with respect to the substrate and product densities.
These condensates were shown to organize the substrate distribution and catalysis-driven fluxes such that the condensates exhibit self-propulsion \cite{Demarchi.etal2023,Goychuk.etal2024}.

Auto-chemotaxis, the directed movement along gradients in chemicals that are secreted by the cells themselves, can lead to the self-organization of cells on the population level via chemotactic aggregation.
Intriguing patterns have been observed, for instance, in experiments with \textit{E.\ coli} \cite{Budrene.Berg1995,Cremer.etal2019}.
The theoretical description of chemotactic aggregation started with the Keller–Segel model to describe accumulation of \textit{Dictyostelium discoideum}~\cite{Keller.Segel1970}.
Since then, many variants of this model have been analyzed, revealing a diverse array of spatiotemporal patterns including aggregate coarsening, stable aggregate sizes, and spatiotemporally chaotic aggregate dynamics.
In particular, the interplay of chemotaxis with cell growth and death yields intriguing patterns \cite{Hillen.Painter2009,Painter.Hillen2011}.
To describe complex biological chemotaxis systems, one must consider more complex signaling dynamics \cite{Ziepke.etal2022} and several cell and chemoattractant species \cite{Wolansky2002,Liu.etal2019a,Muramatsu.etalsubmitted}.

The wide range of different patterns raises the question of how to formulate a theoretical framework and perform a mathematical analysis of the pattern formation process.
Two primary approaches are used to analyze pattern formation described by nonlinear partial differential equations:
amplitude or phase equations and sharp-interface approximations employing singular perturbation theory.

On the one hand, close to a supercritical onset of pattern formation, pattern amplitudes are small, allowing the use of weakly nonlinear analysis to derive the dynamics of long-wavelength variations, known as amplitude equations.
Similarly, long-wavelength modulations of fully nonlinear patterns can be described by phase equations. Remarkably, these amplitude equations are universal, depending solely on the shape of the dispersion relation, symmetries, and conservation laws \cite{Cross.Hohenberg1993}.
It has been shown that the Cahn--Hilliard equation, classically used to describe phase separation \cite{Cahn.Hilliard1958}, is the amplitude equation of an extended Keller--Segel model \cite{Rapp.Zimmermann2019}.
Thus, the universal dynamics of Keller--Segel models close to onset represents phase separation as described by the Cahn--Hilliard equation.

On the other hand, sharp-interface approximations complement the amplitude-equation approach by describing the dynamics of fully nonlinear patterns by the interaction of well-separated, localized, collective degrees of freedom, frequently interfaces and peaks of the pattern~\cite{Ward2006,Wei.Winter2014}.
An analytic treatment of pattern dynamics becomes possible by a singular perturbation theory based on the scale separation between the width of peaks and interfaces $\ell_\mathrm{int}$ and the distance between them, typically the pattern wavelength $\Lambda$.
Other than amplitude equations, these methods retain a description in terms of the physical fields.
This simplifies the interpretation of the quantities arising.
Several model variants of the original Keller--Segel system have been studied using singular perturbation methods for sharp peaks and interfaces~\cite{Ward2006,Hillen.Painter2009}.
The long-time coarsening dynamics was derived for a model including volume filling in Refs.~\cite{Dolak.Schmeiser2005,Potapov.Hillen2005} and for the minimal Keller--Segel model (see below) in Ref.~\cite{Kang.etal2007}.
The effect of cell growth and death was analyzed for the latter model, for instance, in Refs.~\cite{Kolokolnikov.etal2014,Kong.etal2024}.
In addition to these works concerning one-dimensional patterns, sharp-interface flows were derived for pattern interfaces in two dimensions in mesa-forming varients of the Keller--Segel model \cite{Kavanagh2014}.
Two-dimensional patterns under (bistable) cell proliferation have been analyzed in Ref.~\cite{Mimura.Tsujikawa1996} and following publications.

Additionally, coarse-graining particle-based chemotaxis models with linear chemoattractant production or a quasistationary chemoattractant, phase-separation-like behavior has been observed and an effective free energy functional has been constructed \cite{Meyer.etal2014,OByrne.Tailleur2020,Dinelli.etal2024}.
Similarly, Ref.~\cite{Horstmann2001} shows how Lyapunov functionals acting as an effective free energy can be constructed mathematically for Keller--Segel models under conditions on the reaction terms and the diffusion as well as chemotaxis coefficients.
This applies, for instance, if the chemoattractant reactions are linear and the chemotaxis strength is independent of the chemoattractant concentration \cite{Hillen.Painter2009}.

These works suggest that the dynamics without cell growth and death resembles phase separation and aggregates undergo a coarsening process.
However, we are missing a general analysis of Keller--Segel models.
In addition, as the derived instability rates of the periodic patterns are rather complicated, the expressions for specific systems do not show clearly what are the underlying physical processes driving coarsening and which system properties are relevant to the rates.
It is unclear which collective, mesoscopic properties of the aggregates determine the coarsening law and decide whether coarsening occurs dominantly via the competition between aggregates (analogously to Ostwald ripening in phase-separated liquid mixtures) or their coalescence.

Here, we consider the dynamics of chemotactic aggregates in general mass-conserving Keller--Segel models, that is, systems without cell growth and death.
We show that these aggregates generically undergo uninterrupted coarsening by the competition for mass between aggregates and aggregate coalescence.

In Ref.~\cite{Weyer.etalinpreparation}, we have shown that chemotactic aggregates without cell growth and death exactly fulfill a generalized Maxwell construction and argued within a quasi-steady-state (QSS) approximation for the individual aggregates that these systems undergo coarsening.
In this work, we employ singular perturbation theory to complement the QSS analysis by a derivation of the full rates of mass competition between (well-separated) aggregates in mass-conserving Keller--Segel models.
We find a diffusion- and a reaction-limited regime of mass competition similarly as in two-component mass-conserving reaction--diffusion (2cMcRD) systems \cite{Weyer.etal2023} and with the diffusion-limited rate recovering the result obtained by the QSS analysis in Ref.~\cite{Weyer.etalinpreparation}.
In both the diffusion- and reaction-limited regimes, we show that coarsening is driven by self-amplifying mass transport from smaller to larger aggregates and the coalescence of aggregates.

Our results derive the macroscopic dynamics of the Keller--Segel models by introducing a mass-redistribution potential, whose gradients incorporate both diffusion and chemotaxis of the cells and thus drive the redistribution of the cells.
While the mass-redistribution potential for 2cMcRD systems \cite{Brauns.etal2020} is a linear combination of the component densities and pattern formation is driven by a nonlinear reaction term, it is a highly nonlinear function in Keller--Segel models because chemotaxis is described by a nonlinear cross-diffusion term.
We detail the construction of the stationary profiles of the chemotactic aggregates using the mass-redistribution potential, and thereby construct generic expressions quantifying the coarsening of the aggregates.
Our results show that the qualitative macroscopic behavior is unaffected by the different microscopic pattern-forming mechanism, and both 2cMcRD and Keller--Segel systems macroscopically resemble phase-separating mixtures.
Our results generalize the previous mathematical results for specific Keller--Segel models.
Additionally, we provide the linear stability analysis for the lateral instability destabilizing the homogeneous steady state (hss).
We show that the instability is predicted by a nullcline-slope criterion analogously to the curvature criterion for spinodal decomposition.

\textit{Structure of the manuscript.\;---}
We first present the generalized Keller--Segel model that we discuss in the following.
Also, the mass-redistribution potential is introduced, which allows a description of cell motion that combines the counteracting effects of cellular diffusion and chemotactic (directed) movement.
We then introduce the phenomenology of chemotactic aggregation, describing the instability of uniform cell distribution, the formation of the aggregates, and their subsequent coarsening process in Sec.~\ref{sec:phenomenology}.
Afterward, the linear stability analysis of the hss is performed in Sec. \ref{sec:lsa}.
In Sec.~\ref{sec:stat-patterns}, details of the stationary patterns are presented.
Section~\ref{sec:mass-comp} derives the mass-competition rates for peak competition and peak coalescence.
To this end, the assumptions underlying our singular perturbation approach are given and the stability of single half-peaks is discussed in Appendix~\ref{sec:elementary-stab}.
The final results in terms of mesoscopic properties of the aggregates are given in Eqs.~\eqref{eq:sigma-coal}--\eqref{eq:coal-rate-timescale-addition},~\eqref{eq:sigma-comp} and the diffusion- and reaction-limited regimes are discussed.
At the end of the section, the analogous competition rates for mesa patterns are stated in Eqs.~\eqref{eq:sigma-comp-mesa},~\eqref{eq:sigma-coal-mesa}.
Finally, we summarize our results and conclude our discussion in Sec.~\ref{sec:conclusions}.

\section{The Keller--Segel Model}
The Keller--Segel model \cite{Keller.Segel1970,Keller.Segel1971}, as formulated initially, was proposed as a set of four equations that were reduced to a simplified model describing the density of the chemotactic cell population ${\rho(\mathbf{x},t) \geq 0}$ and the chemoattractant ${c(\mathbf{x},t) \geq 0}$ by
\begin{subequations}
\begin{align} 
    \partial_t \rho 
    &= \nabla D_1(\rho,c)\nabla \rho - T \nabla \left[D_2(\rho,c)  \nabla c\right] 
    \, , \\
    \partial_t c 
    &= D_c \nabla^2 c + g(c)\, \rho - h(c) \, c
    \, .
\end{align}
\end{subequations}
The equation for the cell density $\rho$ conserves the total number of the cells and only describes their movement in the system.
The first term describes their random motion by a diffusion term with a (possibly density-dependent) diffusion coefficient $D_1(\rho,c)$.
The second term describes chemotactic, directed movement along gradients $\bm{\nabla}c$ of the chemoattractant.
The chemoattractant spreads diffusively with diffusion coefficient $D_c$, is produced by the cells with a rate $g(c)$, and degraded with a rate $h(c)$.
 
Our discussion concerns the generalized Keller--Segel model (see, for instance, Ref.~\cite{Schaaf1985})
\begin{subequations}\label{eq:KS}
    \begin{align} 
    \partial_t \rho 
    &= D_\rho \nabla^2 \rho - T \nabla  \left[ \rho \chi_\rho(\rho) \chi_c(c) \nabla c \right] 
    \, ,
    \label{eq:rho-KS}\\
    \partial_t c 
    &= D_c \nabla^2 c + f(\rho, c).\label{eq:c-KS}
    \end{align}
\end{subequations}
We assume that the chemoattractant is produced by the cells.
Therefore, we demand $\partial_\rho f > 0$.
This requirement ensures that higher cell densities $\rho$ bias the reaction term toward chemoattractant production.
Also, compared to its initial formulation, we restrict the chemotactic sensitivity to a factorized form $\chi(\rho,c) = \chi_\rho(\rho)\chi_c(c)$ with $\chi_\rho,\chi_c>0$ for all densities $\rho$ and $c$ attained by the patterns.
This simplification has been employed previously when constructing the stationary patterns mathematically \cite{Schaaf1985} and includes a wide range of different model extensions incorporating signal- and density-dependent sensitivities, volume filling, and nonlinear chemoattractant kinetics \cite{Hillen.Painter2009}.
For simplicity, we assume a constant diffusion coefficient $D_\rho$.
The definition of the mass-redistribution potential, the flux-balance, and the reactive area constructions generalize to the case of a diffusion term ${\nabla[D_\rho(\rho)D_c(c)\nabla\rho]}$ under the same conditions ${D_\rho, D_c>0}$ as for the chemotactic sensitivities.
Thus, also the long-time dynamics in this more general setting can be analyzed by extending our theory.

The dynamics Eq.~\eqref{eq:KS} conserve the average cell density
\begin{equation}
    \bar{\rho} = \frac{1}{|\Omega|} \int_\Omega\mathrm{d}^dx\, \rho(\mathbf{x},t).
\end{equation}
It is constant on closed domains and set by the initial condition.
We consider domains with no-flux or periodic boundary conditions.
The effect of broken mass conservation, that is, cell growth and death, is analyzed in Ref.~\cite{Weyer.etalinpreparation}.

We consider the dynamics, Eq.~\eqref{eq:KS},  on a $d$-dimensional domain ${\Omega \subset \mathbb{R}^d}$ with no-flux or periodic boundary conditions.
Most results are obtained for the patterns on a one-dimensional finite interval $I$ with no-flux or periodic boundary conditions, or on the infinite line.
The concepts can be generalized to higher-dimensional systems as discussed for mass-conserving reaction--diffusion systems in Refs.~\cite{Brauns.etal2021,Weyer.etalsubmitted}

We exemplify our general expression using the \emph{minimal Keller--Segel (mKS) model}~\cite{Childress.Percus1981}
\begin{subequations}\label{eq:mKS}
    \begin{align} 
    \partial_t \rho 
    &= D_\rho \nabla^2 \rho - T \nabla  \left[ \rho \nabla c \right]
    \, , \\
    \partial_t c 
    &= D_c \nabla^2 c + \rho - c
    \, .
    \end{align}
\end{subequations}
Here, the chemotactic sensitivities ${\chi_\rho = \chi_c = 1}$ are set constant and the chemoattractant is produced and degraded linearly.

\subsection{The mass-redistribution potential}
In Ref.~\cite{Weyer.etalinpreparation}, we discuss that the equation of motion for the cell density $\rho$, Eq.~\eqref{eq:rho-KS}, can be rewritten in terms of a mass-redistribution potential (cf.\ Refs.~\cite{Otsuji.etal2007,Halatek.Frey2018,Brauns.etal2020} for its role in mass-conserving reaction--diffusion systems and Ref.~\cite{Cotton.etal2022} for its definition in a related cross-diffusion system), the gradients of which drive the redistribution of the conserved cell density within the system.
We discuss its behavior in more detail here.
Equation~\eqref{eq:rho-KS} gives
\begin{align} 
\partial_t \rho &= \nabla \left[ T \chi_\rho \rho \left( \frac{D_\rho}{T \chi_\rho \rho}\nabla\rho - \chi_c \nabla c \right)\right] \nonumber\\
&= \nabla \left[ T \chi_\rho \rho \nabla \eta\right] .
\end{align}
The last equality defines the mass redistribution potential
\begin{equation} 
\label{eq:SI_eta_def}
    \eta 
    \equiv 
    \frac{D_\rho}{T} \, \eta_\rho - \eta_c
    \, , 
\end{equation}
with
\begin{subequations}
\begin{align} 
    \eta_\rho(\rho) 
    &\equiv \int_{\rho_0}^{\rho} \mathrm{d}\rho'\, \frac{1}{\chi_\rho(\rho') \rho'}
    \, , \\  
    \eta_c(c)
    &\equiv \int_{c_0}^{c} \mathrm{d}c'\, \chi_c(c')
    \, ,
\end{align} 
\end{subequations}
choosing the arbitrary reference densities $\rho_0$ and $c_0$.

The definition of the mass-redistribution potential Eq.~\eqref{eq:SI_eta_def} results in the continuity equation for the cell density 
\begin{equation}
    \label{eq:cont-eq}
    \partial_t \rho = \nabla \left[ T \rho \chi_\rho \nabla \eta\right].
\end{equation}
Thus, the gradients in the mass-redistribution potential drive the redistribution of the cells.
Specifically, the flux $\mathbf{J}$ of the cell density is given by ${\mathbf{J} = - T \rho \chi_\rho \bm{\nabla} \eta}$, i.e., cells migrate from regions with high values of the mass-redistribution potential to regions with lower values of the mass-redistribution potential.
Comparison with the original Eq.~\eqref{eq:KS} shows that the mass-redistribution potential $\eta$ can be interpreted as an effective chemorepellent for $\rho$ that encompasses the effect of cellular diffusion.
Thus, the counteracting effects of random (diffusive) and directed (chemotactic) cell motion is described in terms of the single potential $\eta$ instead of two independent flux terms ${-D_\rho \bm{\nabla} \rho}$ and ${T \rho \chi_\rho \chi_c \bm{\nabla} c}$.
The increase of $\eta$ with $\rho$ describes the diffusive redistribution from high to low cell density while the decrease of $\eta$ with $c$ captures the aggregation due to attractive auto-chemotaxis.

Moreover, the mass-redistribution potential is similar to the chemical potential in Model B dynamics \cite{Hohenberg.Halperin1977,Bray2002}.
In contrast to a chemical potential, the mass-redistribution potential is governed by its own equation of motion.
The definition of $\eta$, Eq.~\eqref{eq:SI_eta_def}, together with the generalized Keller--Segel equations Eqs.~\eqref{eq:KS} yields
\begin{equation}
\partial_t \eta = \frac{D_\rho}{T} \frac{1}{\chi_\rho \rho} \partial_t \rho - \chi_c \partial_t c\, ,
\end{equation}
and hence,
\begin{widetext}
\begin{align}
\partial_t \eta &= \frac{D_\rho}{T} \frac{\partial_t \rho}{\chi_\rho \rho} + D_c\chi_c\nabla \frac{1}{\chi_c}\nabla \eta - \chi_c\frac{D_\rho D_c}{T}\nabla \frac{1}{\chi_c} \nabla\eta_\rho - \chi_c f(\rho_, c)\, , \nonumber\\
\label{eq:KS-eta}
&= \frac{D_\rho}{\chi_\rho \rho} \nabla \chi_\rho \rho \nabla \eta + D_c \chi_c \nabla \frac{1}{\chi_c} \nabla \eta - \chi_c \frac{D_\rho D_c}{T} \nabla \frac{1}{\chi_c \chi_\rho \rho} \nabla \rho - \chi_c \Tilde{\Tilde{f}}(\rho, \eta)\, . 
\end{align}
\end{widetext}
The first two terms describe the (generalized) diffusion of the mass-redistribution potential.
The second two terms are source terms that depend on the cell-density profile $\rho$.

Because we assume $\chi_\rho,\chi_c>0$ and $\rho,c\geq 0$, the functions $\eta_\rho(\rho),\eta_c(c)$ are invertible.
Thus, one can choose any two of the three variables $\rho$, $c$, and $\eta$ as independent and express the third one as a function of the others:
\begin{subequations}\label{eq:density-expressions}
    \begin{align}
        \eta(\rho,c) &= \frac{D_\rho}{T} \eta_\rho - \eta_c,\\
        c(\rho, \eta) &= \eta_c^{-1}\left(\frac{D_\rho}{T} \eta_\rho-\eta\right),\label{eq:c-expression}\\
        \rho(c, \eta) &= \eta_\rho^{-1}\left(\frac{T}{D_\rho} (\eta+\eta_c)\right).\label{eq:rho-expression}
    \end{align}
\end{subequations}
With this, we also define the short-hand notations for the reaction term
\begin{subequations}\label{eq:f-expressions}
\begin{align}
    \Tilde{f}(c,\eta) 
    &\equiv f(\rho(c,\eta),c)
    \, ,\\
    \Tilde{\Tilde{f}}(\rho,\eta) 
    &\equiv f(\rho,c(\rho,\eta))
    \, .
\end{align}
\end{subequations}

To illustrate the mass-redistribution potential, consider the mKS model.
In this model, one has
\begin{equation}
    \eta_\rho = \log \rho\, , \qquad \eta_c = c\, ,
\end{equation}
where we dropped constants arising from the integration boundaries $\rho_0$ and $c_0$.
This gives
\begin{subequations}
\begin{align}
    \eta(\rho,c) &= \frac{D_\rho}{T}\log \rho - c\, ,\label{eq:mKS-eta}\\
    c(\rho,\eta) &= \frac{D_\rho}{T}\log \rho - \eta\, ,\\
    \rho(c,\eta) &= \exp\left[\frac{T}{D_\rho}(\eta+c)\right].
\end{align}
\end{subequations}
The mKS dynamics then read
\begin{subequations}
    \begin{align}
        \partial_t \rho &= \nabla\left[ T \rho \nabla \eta\right]
        \, , \\
        \partial_t \eta 
        &= \frac{D_\rho}{\rho} \nabla\left[ \rho \nabla\eta\right] + D_c \nabla^2 \eta 
        \nonumber \\ 
        &\quad - \frac{D_\rho D_c}{T} \nabla\left[ \frac{1}{\rho}\nabla\rho\right] - \Tilde{\Tilde{f}}(\rho,\eta)
        \, .
    \end{align}
\end{subequations}
The dynamics of the mass-redistribution potential contains (modified) diffusion terms deriving from the redistribution of the cell and chemoattractant densities.
The last two terms correspond to changes in $\eta$ that depends on the cell density profile.
These terms turn out to play a role analogous to the functional relation for the chemical potential, ${\mu = \delta \mathcal{F}/\delta\phi}$, in binary phase-separation dynamics with free energy functional $\mathcal{F}[\phi]$ and the volume fraction $\phi$ (see Sec.~\ref{sec:stat-patterns} and Refs.~\cite{Weyer.etalinpreparation,Weyer.etal2023}).

\subsection{Reactive equilibria}
\label{sec:local-dynamics}
In each sufficiently small subregion of the system, the densities are approximately constant.
Within the subregion, the dynamics correspond to the dynamics of the well-mixed system (``local dynamics'')
\begin{subequations}\label{eq:local-dyn}
    \begin{align}
        \partial_t \rho &= 0,\\
        \partial_t c &= f(\rho,c).
    \end{align}
\end{subequations}
Thus, while the reactions $f$ act locally, the diffusive and chemotactic mass transport of the full system Eq.~\eqref{eq:KS} couples each local subregion to neighboring regions.

Because the Keller--Segel dynamics Eq.~\eqref{eq:KS} is mass-conserving, the cell density remains constant in the well-mixed system, set by the initial condition.
In contrast, the reactions drive the chemoattractant density to equilibrate to the reactive equilibrium $c^*(\rho)$ defined by
\begin{equation}
    f(\bar\rho,c^*(\rho))=0.
\end{equation}
Equivalently, the mass-redistribution potential $\eta$ follows the local dynamics 
\begin{equation}
    \partial_t \eta 
    = - \chi_c \Tilde{\Tilde{f}}(\bar\rho,\eta)
\end{equation} 
towards the reactive equilibrium $\eta^*(\rho)$.
The family of reactive equilibria $[\eta^*(\rho), c^*(\rho)]$ for different densities $\rho$ is called the nullcline (NC).
If the reactions exhibit multistability, several reactive equilibria exist for a fixed density $\rho$.
Multistable reaction kinetics can be treated in $(c,\eta)$ phase space, as described in the Supplemental Material to Ref.~\cite{Weyer.etalinpreparation}. 

\begin{figure*}
    \centering
    \includegraphics[]{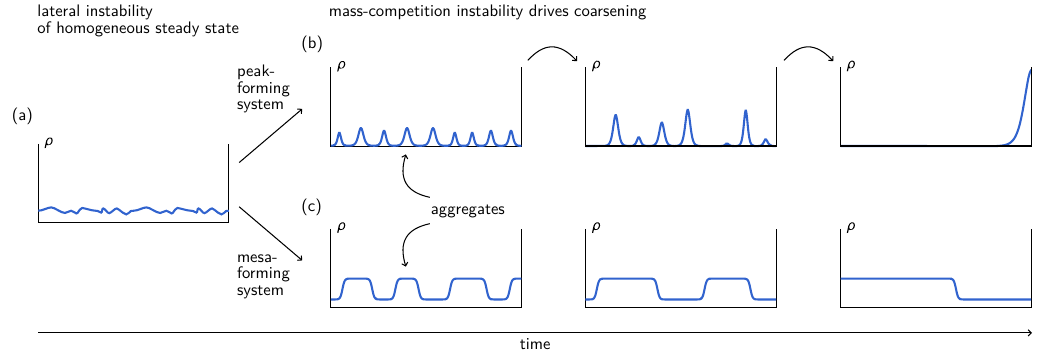}
    \caption{The phenomenology of chemotactic aggregation.
    (a) If chemotaxis is sufficiently strong (see Sec.~\ref{sec:lsa}), a uniform distribution of the cells is unstable against spatially varying perturbations (lateral instability).
    The result is the aggregation of cells into aggregates that are either peak- (b) or mesa-shaped (c).
    These aggregates are quasi-stationary but interact leading to a coarsening of the aggregates by competition for cells and coalescence (mass-competition instability, see Sec.~\ref{sec:mass-comp}).
    The final stationary pattern contains a single aggregate that moves to the system boundary in the case of no-flux boundary conditions of the domain (illustrated here).
    }
    \label{fig:phenomenology}
\end{figure*}

\section{Phenomenology of chemotactic aggregation}
\label{sec:phenomenology}
Before we begin the detailed analysis, we will present the generic phenomenology of the chemotactic aggregates and their dynamics that we derive in the following.

In the Keller--Segel models we consider, the cells produce a chemoattractant and perform chemotaxis along gradients in its density (auto-chemotaxis).
Thus, the chemoattractant mediates an effective attraction between the cells.
If this process overcomes the random movement of the cells (diffusion), a uniform distribution of the cells is unstable against weak perturbations and cells accumulate into high-density aggregates separated by low-density regions [see Fig.~\ref{fig:phenomenology}(a), Sec.~\ref{sec:lsa}].
Two types of (well-separated) aggregates can be distinguished (see Sec.~\ref{sec:stat-patterns}): peak and mesa patterns [see Fig.~\ref{fig:phenomenology}(b,c)].
Mesa-shaped aggregates form if the cell density saturates and a high-density plateau forms.
In contrast, peak-shaped aggregates result if the cell density does not saturate, for instance, because the total mass of the aggregate is so low that the maximal density in the aggregates lies below the saturation density.
Peak patterns form in the mKS model, while mesa patterns arise, for example, in the Keller--Segel model with volume filling \cite{Hillen.Painter2001,Painter.Hillen2002}.

In a large system, several quasi-stationary aggregates form initially.
These different aggregates will have slightly different total masses.
Using the generalized Maxwell construction (see Sec.~\ref{sec:stat-patterns} and Ref.~\cite{Weyer.etalinpreparation}), we show that larger aggregates attract cells more strongly, and thus, deplete the surrounding low-density plateau more strongly of cells than smaller aggregates \cite{Weyer.etalinpreparation}.
As a result, larger aggregates slowly accumulate more and more cells at the expense of smaller aggregates which loose cells into the low-density plateaus until they collapse.
Consequently, this slow dynamics results in a coarsening process of the pattern that reduces the number of aggregates until a single aggregate remains.
Similarly to this competition process, two aggregates attract each other, an effect that is stronger if the aggregates are closer to each other.
As a result, the aggregates also undergo coalescence.
We describe these processes analytically for peak patterns and state the results for peak patterns in Sec.~\ref{sec:mass-comp}.
These results, together with the generalized Maxwell construction (see Sec.~\ref{sec:stat-patterns}) for the stationary aggregates, imply that coarsening is generically uninterrupted and proceeds until all cells accumulate into a single aggregate [see Fig.~\ref{fig:phenomenology}(b,c)].

\section{Linear stability analysis of the homogeneous steady state}
\label{sec:lsa}
In Ref.~\cite{Weyer.etalinpreparation}, we have argued via a QSS approximation that the growth of spatial variations around the homogeneous steady state $[\rho_\mathrm{hss},\eta^*(\rho_\mathrm{hss})]$ (lateral instability of the homogeneous steady state while assuming that the local dynamics is stable) in the Keller--Segel dynamics can be explained by the feedback arising from the decrease of the local steady-state mass-redistribution potential $\eta^*(\rho)$ with increasing density $\rho$.
Thus, the resulting instability condition is the \emph{nullcline-slope condition}
\begin{equation} 
\label{eq:SI_steady_state_approx_lin_instab}
\partial_\rho \eta^*\big|_{\rho_\mathrm{hss}} < 0\, .
\end{equation}
This is the same criterion as the curvature criterion (thermodynamic stability of the spatially uniform state) for the free energy in phase-separating systems~\cite{Doi2013}. 
Given a perturbation ${\delta\rho \sim \operatorname{e}^{i \mathbf{q}\cdot\mathbf{x}}}$, the QSS approximation applies in the long-wavelength limit ${q=|\mathbf{q}|\to 0}$ because in this limit, the redistribution of mass between the high- and low-density regions becomes slow compared to the local reactive dynamics Eq.~\eqref{eq:local-dyn}.

What is the interpretation of this slope criterion?
Recall that the mass-redistribution potential increases with the cell density $\rho$, capturing diffusion from high- to low-density regions, and decreases with the chemoattractant density $c$, describing chemotaxis from regions of low to regions of high chemoattractant density.
The nullcline-slope criterion thus requires that the local steady-state chemoattractant density increases with the cell density so strongly that chemotaxis towards these high-density regions overcomes diffusive spreading.
Consequently, if the slope-criterion is fulfilled, perturbations in the cell density self-amplify due to chemotactic aggregation.

In order to show that the QSS argument is exact for arbitrary wavenumbers $q$, we now calculate the full dispersion relation and demonstrate that the lateral instability of Eqs.~\eqref{eq:KS} is always of type II (long-wavelength instability) under the classification scheme of Cross and Hohenberg \cite{Cross.Hohenberg1993}.
In a type-II unstable system, long wavelength perturbations with wavenumbers $q\to 0$ become unstable first.
Thus, it is sufficient to consider the stability in the QSS limit $q\rightarrow0$.

Equations~\eqref{eq:KS} can be written in the compact form 
\begin{equation} 
\label{eq:SI-eom-vector}
    \partial_t \mathbf{u} = \nabla \left[\mathbf{M}(\mathbf{u})\nabla\mathbf{u}\right] + \mathbf{f}(\mathbf{u})\, ,
\end{equation}
where ${\mathbf{u} := (\rho, c)^\mathsf{T}}$ encompasses both $\rho$ and $c$, the reaction term ${\mathbf{f}(\mathbf{u}) = [0, f(\rho, c)]^\mathsf{T}}$ contains the chemokine reaction dynamics, and the mobility matrix $\mathbf{M} (\mathbf{u})$ contains both diffusion and chemotaxis terms, giving
\begin{equation} 
\mathbf{M}(\mathbf{u}) = \begin{pmatrix}
    D_\rho & - T \chi_\rho \chi_c \rho \\
    0      & D_c
\end{pmatrix}.
\end{equation}
The homogeneous steady state $\mathbf{u}_\mathrm{hss}$ solves ${0 = \mathbf{f}(\mathbf{u}_\mathrm{hss})}$ (cf.~Sec.~\ref{sec:local-dynamics}).
To analyze the stability of the uniform steady state $\mathbf{u}_\mathrm{hss}$ against small perturbations, we linearize Eqs.~\eqref{eq:KS} around $\mathbf{u}_\mathrm{hss}$ in ${\delta\mathbf{u} = \mathbf{u}-\mathbf{u}_\mathrm{hss}}$, resulting in
\begin{equation} 
\partial_t\delta\mathbf{u} 
= [\mathbf{M}(\mathbf{u}_\mathrm{hss})\nabla^2 + \mathbf{J}]\delta\mathbf{u}\, ,
\end{equation}
with the components of the Jacobian of the reaction terms
\begin{equation} 
J_{ij} = \left.\partial_{u_j} f_i\right|_{\mathbf{u} = \mathbf{u}_\mathrm{hss}} .
\end{equation}
To determine the stability of the homogeneous steady state, one determines the eigenmodes and eigenvalues.
If some eigenvalue $\sigma_q$ has a positive real part, this eigenmode will grow exponentially over time, indicating an instability.
Choosing the ansatz of an eigenmode expansion of the form
\begin{equation} 
    \delta \mathbf{u} (\mathbf{x}, t) = \sum_{\mathbf{q}} e^{\sigma_{q} t} e^{i \mathbf{q} \cdot \mathbf{x}}\delta \mathbf{u}_{\mathbf{q}}\, ,
\end{equation}
where the sum spans all wavevectors $\mathbf{q}$,
we obtain an equation for the growth rates $\sigma_q$ for the Fourier modes with wavenumber $q=|\mathbf{q}|$ 
\begin{equation} 
    \sigma_{q} \delta \mathbf{u}_{\mathbf{q}} = \left[-q^2 \mathbf{M}(\mathbf{u}_\mathrm{HSS}) + \mathbf{J}\right]\delta \mathbf{u}_{\mathbf{q}}\, .
\end{equation}
As the dynamic matrix is two-dimensional, the dispersion relation has the two branches
\begin{widetext}
\begin{equation} \label{eq:full-disp-rel}
    \sigma_{q}^\pm 
    = 
    \frac{1}{2}\left(-\big[(D_\rho + D_c) q^2 - \partial_c f \big] \pm \sqrt{
    \big[ (D_\rho + D_c) q^2 - \partial_c f \big]^2 - 4 \big[D_\rho q^2(D_c q^2 - \partial_c f) - (\partial_\rho f) T \rho_\mathrm{hss} \chi_c\chi_\rho q^2 \big]}\right),
\end{equation}
\end{widetext}
with all expressions evaluated at the homogeneous steady state densities $(\rho_\mathrm{hss},c_\mathrm{hss})$.
Setting ${q = 0}$, one finds (cf.~Sec.~\ref{sec:local-dynamics}) that the local stability of the homogeneous steady state requires that ${\partial_c f < 0}$.
The mass-conservation law implies that one eigenvalue of the local dynamics is zero.
The corresponding eigenvector is the tangent vector of the nullcline.

\begin{figure}[tb]
    \centering
    \includegraphics[]{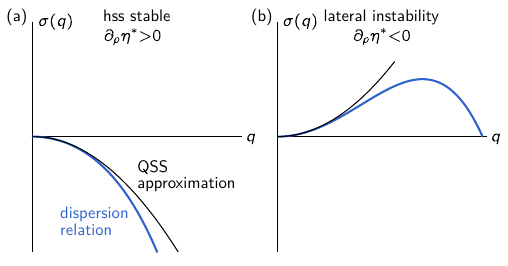}
    \caption{Dispersion relation for the lateral instability of the hss.
    (a) If the nullcline slope is negative, all growth rates $\sigma_q^+$ are negative (blue), as also the QSS approximation of the dispersion relation for $q\to0$ (black).
    (b) A band of unstable modes that extends to $q=0$ emerges if the nullcline slope is negative.
    }
    \label{fig:dispRel}
\end{figure}

The real parts of the eigenvalues $\sigma_q^-$ are negative and the lateral stability is solely determined by the branch $\sigma_q^+$.
Moreover, the real part only becomes positive if the square root is positive, such that $\sigma_q^+$ will be real.
Consequently, the band of unstable modes is bounded by the wavenumbers $q$ at which $\sigma_q^+=0$.
Solving for $q\geq0$ with $\sigma_q^+ = 0$ shows that $\sigma_q^+$ has only two roots, 
\begin{equation} 
q_0 = 0\, , \qquad q_1 = \sqrt{\frac{1}{D_c}\left(\frac{(\partial_\rho f) T}{D_\rho} \rho_\mathrm{hss}\chi_\rho \chi_c + \partial_c f\right)}\, ,
\end{equation}
and $\sigma_q^+>0$ for $q_0< q< q_1$.
Thus, the band of unstable modes always extends to $q=0$ and the instability is of type II (see Fig.~\ref{fig:dispRel}).

As a result, stability can be determined in the long-wavelength limit $q\to 0$.
Expanding the dispersion relation Eq.~\eqref{eq:full-disp-rel} in $q$ then yields 
\begin{equation} 
\label{eq:expanded_eigenvals}
\sigma_q^+ = -\left(D_\rho + \frac{\partial_\rho f}{\partial_c f} T \rho_\mathrm{hss} \chi_\rho\chi_c\right) q^2 + \mathcal{O}(q^4)\, .
\end{equation}
As stated in Sec.~\ref{sec:local-dynamics}, the nullcline $\eta^*(\rho)$ is determined by ${f (\rho, c^*) = \Tilde{\Tilde{f}}(\rho, \eta^*) = 0}$.
Employing the implicit function theorem, one obtains
\begin{equation} 
\partial_\rho \eta^* = \frac{D_\rho}{T \chi_\rho \rho} + \chi_c \frac{\partial_\rho f}{\partial_c f}\, ,
\end{equation}
which gives
\begin{equation}\label{eq:long-wavelength-eigenvalue}
\sigma_q^+ = -T \chi_\rho \rho_\mathrm{hss}(\partial_\rho \eta^*|_{\rho_\mathrm{hss}}) q^2 + \mathcal{O}(q^4).
\end{equation}

Within the QSS approximation based on approximating the local densities by the reactive equilibria $\Tilde{\Tilde{f}}=0$, i.e., setting $\eta\approx \eta^*(\rho)$, the redistribution dynamics of the conserved density $\rho$, that is, the continuity equation Eq.~\eqref{eq:cont-eq} becomes
\begin{equation}
    \partial_t \rho \approx \nabla\left[ T \chi_\rho \rho \nabla \eta^*(\rho)\right].
\end{equation}
Expanding to first order in variations $\delta\rho$ around the homogeneous steady state $\rho_\mathrm{hss}$, one has
\begin{equation}
    \partial_t \delta\rho \approx T \chi_\rho(\rho_\mathrm{hss}) \rho_\mathrm{hss} \partial_\rho \eta^*|_{\rho_\mathrm{hss}}\nabla^2\delta\rho\, ,
\end{equation}
which gives the eigenvalue
\begin{equation}
    \sigma_\mathrm{qss} = -T \chi_\rho(\rho_\mathrm{hss}) \rho_\mathrm{hss} \partial_\rho \eta^*|_{\rho_\mathrm{hss}} q^2\, ,
\end{equation}
in agreement with Eq.~\eqref{eq:long-wavelength-eigenvalue} (see Fig.~\ref{fig:dispRel}).

Thus, a lateral instability due to an effective anti-diffusion of the cell density $\rho$, called a \emph{mass-redistribution instability} \cite{Brauns.etal2020}, occurs indeed if (cf.\ Ref.~\cite{Weyer.etalinpreparation})
\begin{equation}\label{eq:nullcline-slope-crit}
    \partial_\rho \eta^*|_{\rho_\mathrm{hss}} < 0\, .
\end{equation}
This nullcline-slope criterion is exact arbitrarily far from the long-wavelength (QSS) limit.

\section{Stationary patterns}
\label{sec:stat-patterns}
In this section, we provide a detailed analysis of the one-dimensional (periodic) stationary patterns that emerge from the Keller--Segel dynamics given by Eqs.~\eqref{eq:KS}.
The one-dimensional pattern profiles can be used to construct stationary patterns in higher-dimensional systems in the limit of weakly curved interfaces, as described in Refs.~\cite{Bray2002,Pismen2006,Brauns.etal2021,Weyer.etalsubmitted}.
Moreover, we note that our analysis of the mass-competition instability in Sec.~\ref{sec:mass-comp} will show that the periodic patterns are unstable against mass transport between and the coalescence of aggregates.
Only the stationary patterns with a single aggregate are stable [situated at the boundary in the case of no-flux boundary conditions, see right-most panels in Fig.~\ref{fig:phenomenology}(b,c)].

In steady state, the density flux ${J=- T\chi_\rho \rhoStat \partial_x \etaStat}$ in the continuity equation Eq.~\eqref{eq:cont-eq} must be spatially uniform.
As we assume $\chi_\rho,\rho\geq 0$, a finite flux $J$ would imply a difference between the values of mass-redistribution potential at the domain boundaries, $\etaStat(0)$ and $\etaStat(L)$.
Thus, on one-dimensional domains with either no-flux or periodic boundary conditions, as well as on an infinite line, the flux must vanish, implying
\begin{equation}
    \etaStat = \mathrm{const.}
    \, ,
\end{equation}
a condition analogous to a constant chemical potential in thermodynamics.
In the $(\rho,\eta)$ phase space, we refer to this line as the flux-balance subspace (FBS)~\cite{Brauns.etal2020}.
Physically, the uniform mass-redistribution potential implies that the overall flux of the cells vanishes, that is, the diffusive flux spreading of the aggregate is exactly balanced by the chemotactic aggregation flux.

Using that the continuity equation requires the stationary mass-redistribution potential to be uniform, the equation of motion for the mass-redistribution potential, Eq.~\eqref{eq:KS-eta}, determines the stationary density profile $\rhoStat(x)$ by (\emph{profile equation})
\begin{equation}\label{eq:profile-eq}
    0 
    = 
    \frac{D_\rho D_c}{T}\partial_x \left[\frac{1}{\chi_c \chi_\rho \rhoStat} \, \partial_x \rhoStat\right] 
    + 
    \Tilde{\Tilde{f}}(\rhoStat, \etaStat)
    \, .
\end{equation}
Defining the uniform mass-redistribution potential $\etaStat$, the stationary pattern profile can be determined from this single equation instead of the coupled equations for $\rhoStat$ and $\cStat$.
This approach was introduced in Ref.~\cite{Schaaf1985}, defining the stationary mass-redistribution potential mathematically as an integration constant.

Because the Keller--Segel dynamics is translationally invariant and parity symmetric, one can construct the periodic patterns by reflecting and concatenating \emph{elementary patterns} which comprise half a wavelength $\Lambda$ of the periodic pattern [cf.\ right-most panels in Fig.~\ref{fig:phenomenology}(b,c)].
In the following, we construct these elementary patterns on a domain of length ${L = \Lambda/2}$ with no-flux boundary conditions, which then give rise to periodic patterns of wavelength $\Lambda$.

In the profile equation, Eq.~\eqref{eq:profile-eq}, the stationary mass-redistribution potential $\etaStat$ is a parameter, such that one can parameterize the stationary elementary profile as
\begin{equation}
    \rhoStat(x) = \rhoStat(x;\etaStat)\, .
\end{equation}
As we will see explicitly below, changing the value of $\etaStat$ corresponds to changing the system's average cell density $\Bar{\rho}$.
The average cell density is a control parameter of the stationary pattern because it is set by the initial condition.
Consequently, the change of the stationary profile when mass is added or taken out of the system is given by
\begin{equation}
    \partial_{\Bar{\rho}} \rhoStat(x) 
    = 
    \big[ \partial_{\Bar{\rho}}\etaStat (\Bar{\rho}) \big] \, 
    \partial_{\etaStat} \rhoStat(x)
    \, .
\end{equation}
We therefore call $\partial_{\etaStat} \rhoStat(x)$ the mass mode.
It is an eigenmode of the linear dynamics around the stationary pattern $\rhoStat(x)$.
Specifically, it is a zero mode because it leads from one stationary state to another.

Moreover, because any two of the three variables $\rho$, $c$, and $\eta$ define the third, determining the stationary pattern from the profile equation, Eq.~\eqref{eq:profile-eq}, is equivalent to finding the stationary profile $c_\mathrm{stat}(x;\etaStat)$ from the stationary solution of Eq.~\eqref{eq:c-KS} (\emph{chemoattractant-profile equation})
\begin{equation}\label{eq:profile-eq-c}
    0 
    = 
    D_c \partial_x^2 c_\mathrm{stat} 
    + 
    \Tilde{f}(c_\mathrm{stat},\etaStat)
    \, .
\end{equation}
Indeed, one has from Eq.~\eqref{eq:c-expression}
\begin{equation}\label{eq:c-rho-stat-relation}
    \partial_x \cStat 
    = 
    \frac{D_\rho}{T\chi_c\chi_\rho \rhoStat} \, 
    \partial_x \rhoStat
    \, ,
\end{equation}
that is, Eq.~\eqref{eq:profile-eq-c} is merely a reparametrization of the profile equation for $\rhoStat$, Eq.~\eqref{eq:profile-eq}.
From Eq.~\eqref{eq:c-rho-stat-relation} it also follows that the chemoattractant density $\cStat$ increases monotonously with the cell density $\rhoStat$.
In the following, we will use whichever of the profile equations, Eq.~\eqref{eq:profile-eq} or Eq.~\eqref{eq:profile-eq-c}, is more convenient.

\begin{figure*}
    \centering
    \includegraphics[]{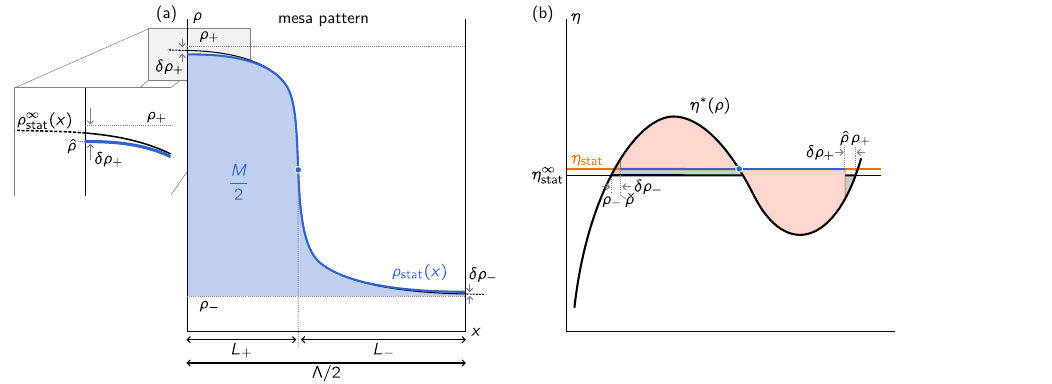}
    \caption{Construction of the stationary elementary mesa pattern.
    (a) The elementary mesa pattern (blue) consists of a single interface separating a high- and a low-density plateau on a domain of length $\Lambda/2$ with no-flux boundary conditions.
    The high- and low-density plateau lengths $L_\pm$ depend on the mesa mass $M$ (blue-shaded region) and, equivalently, the average cell density $\bar{\rho}$ in the system.
    The profile of a single interface on the infinite line (black profile) approaches the plateau densities $\rho_\pm$ exponentially far from the interface.
    The profile on the finite interval of length $\Lambda/2$ (blue) shows deviations in the plateaus close to the boundaries as it must fulfill the boundary conditions.
    As a result, the pattern maximum $\hat{\rho} = \rho_+-\delta\rho_+$ differs from the plateau density $\rho_+$ by an amount $\delta\rho$ exponentially small in the plateau lengths.
    Similarly, the pattern minimum is $\check{\rho}=\rho_-+\delta\rho_-$.
    (b) In the local $(\rho,\eta)$ phase space, the stationary pattern is restricted to the flux-balance subspace (FBS) $\etaStat=\mathrm{const.}$
    The value of $\etaStat$ is fixed qualitatively by the balance of the red-shaded areas between the FBS and the nullcline $\eta^*(\rho)$ (see Ref.~\cite{Weyer.etalinpreparation}).
    The plateau densities $\rho_\pm$ are the intersections of the FBS $\etaStat^\infty$ of the pattern on the infinite line with the nullcline, which have a positive nullcline slope.
    The middle intersection with a negative nullcline slope (blue dot) indicates the inflection point of the pattern profile.
    The shift $\etaStat-\etaStat^\infty$ of the stationary mass-redistribution potential $\etaStat$ of the pattern on a finite domain compared to its value $\etaStat^\infty$ for the pattern on the infinite line is due to the changes of the red-shaded compared to the gray-shaded areas induced by the deviations $\delta\rho_\pm$ in the plateaus.
    }
    \label{fig:stat-mesa}
\end{figure*}

The value of the stationary mass-redistribution potential $\etaStat$ is related to the pattern profile by the constraint that the profile equation, Eq.~\eqref{eq:profile-eq} or Eq.~\eqref{eq:profile-eq-c}, must have a solution that fulfills the boundary conditions.
No-flux or periodic boundary conditions for the chemoattractant imply that, in steady state, the total chemoattractant production must balance the total degradation within the domain.
This constraint follows from integrating Eq.~\eqref{eq:profile-eq-c} over the whole domain which yields
\begin{equation}
    0 = \int_0^\frac{\Lambda}{2}\mathrm{d}x\, \Tilde{f}(\cStat,\etaStat)\, .
\end{equation}
Alternatively, multiplying Eq.~\eqref{eq:profile-eq-c} first by $\partial_x \cStat$, one derives the \emph{total turnover balance} condition (cf.~Ref.~\cite{Brauns.etal2020} for the analogous condition in two-component mass-conserving reaction--diffusion systems)
\begin{equation}\label{eq:ttb}
    0 = \int_{\check{c}}^{\hat{c}}\mathrm{d}c\, \Tilde{f}(c,\etaStat) = \int_{\check{\rho}}^{\hat\rho}\mathrm{d}\rho \frac{\Tilde{\Tilde{f}}(\rho, \etaStat)}{\chi_c\chi_\rho \rho}\, ,
\end{equation}
where the second equality uses the reparametrization from $c$ to $\rho$.
Restricting the construction to monotonous elementary patterns, the densities $\check{\rho},\check{c}$ and $\hat{\rho},\hat{c}$ are the minimal and maximal cell and chemoattractant densities.\footnote{%
At a pattern maximum or minimum, both $\partial_x\rhoStat$ and $\partial_x\cStat$ vanish [cf.~Eq.~\eqref{eq:c-rho-stat-relation}] such that the pattern can be split into monotonous elementary patterns on separate domains with no-flux boundary conditions.
}
In Ref.~\cite{Weyer.etalinpreparation}, we show that this total turnover balance corresponds to a reactive area balance in phase space (see Figs.~\ref{fig:stat-mesa},~\ref{fig:stat-peak}).
This construction leads to generalized Maxwell construction fulfilled by stationary aggregates.
In short, the local reactive turnover vanishes at the chemical equilibria, i.e., the nullcline $\eta^*(\rho)$.
Thus, the distance from the nullcline is related to the strength of the local reactive turnover (close to the nullcline), and the (signed) area enclosed between FBS and NC is a measure for the total turnover, discussed in detail in Ref.~\cite{Weyer.etalinpreparation}.
In the part of the pattern lying above the NC in phase space, the chemoattractant is produced.
In the part of the pattern falling below the NC in phase space, the chemoattractant is degraded.
The balance of the respective two areas between FBS and NC corresponds to the balance of total production and degradation of the chemoattractant along the pattern.
We now discuss general properties of mesa and peak patterns that will be relevant to mass competition in systems containing several elementary patterns.

\subsection{Elementary mesa patterns}
The stationary interface profile of mesa patterns can be discussed by considering a single interface on the infinite line.
In the plateau regions, small corrections to the stationary pattern occur on a finite domain.
These small deviations drive the coarsening process as we show below.

\subsubsection{Stationary interface profile on the infinite line}
Moving the domain boundaries to infinity, we consider an elementary pattern consisting of an interface (positioned around ${x = 0}$) separating an infinite low- and an infinite high-density plateau (cf.~Ref.~\cite{Brauns.etal2021} for the analogous discussion for two-component mass-conserving reaction--diffusion systems).
Far from the interface, the profile curvature vanishes
\begin{equation}
    \lim_{x\to\pm \infty}\partial_x^2 \cStat
    =
    \lim_{x\to\pm \infty}\partial_x^2 \rhoStat 
    =  0 
    \, .
\end{equation}
Thus, the profile equations for the cell density, Eq.~\eqref{eq:profile-eq}, and for the chemoattractant, Eq.~\eqref{eq:profile-eq-c}, show that the plateau densities approach the FBS-NC intersection points ${c_\pm = c_\pm(\etaStat^\infty)}$ and ${\rho_\pm = \rho_\pm(\etaStat^\infty)}$ (see Fig.~\ref{fig:stat-mesa}).
Here, we defined the value of the stationary mass-redistribution potential on the infinite line $\etaStat^\infty$,
which is fixed (independently of the pattern profile) by the total turnover balance [Eq.~\eqref{eq:ttb}]
\begin{equation}\label{eq:ttb-mesa-infLine}
    0 
    = 
    \int_{c_-(\etaStat^\infty)}^{c_+(\etaStat^\infty)}\mathrm{d}c\, \Tilde{f}(c,\etaStat^\infty)
    \, .
\end{equation}

In the plateau regions, the pattern profile can be determined by linearizing the profile equations, Eqs.~\eqref{eq:profile-eq},~\eqref{eq:profile-eq-c}, around the plateau densities, setting ${\rhoStat(x) = \rho_\pm \mp \delta\rho_\pm^\infty(x)}$ [${\cStat(x) = c_\pm \mp \delta c_\pm^\infty(x)}$].
This yields
\begin{align}
    0 &= \frac{D_\rho D_c}{T \chi_c^\pm\chi_\rho^\pm \rho_\pm} \partial_x^2 \delta\rho_\pm^\infty + \partial_\rho \tilde{\tilde{f}}|_{\rho_\pm} \delta \rho_\pm^\infty\\
    &= D_c \partial_x^2 \delta \rho_\pm^\infty + \partial_c \tilde{f}|_{c_\pm} \delta \rho_\pm^\infty\, ,\label{eq:lin-profile-eq}
\end{align}
and an equivalent equation for $\delta c_\pm$.
Here, we defined ${\chi_\rho^\pm = \chi_\rho(\rho_\pm)}$ and ${\chi_c^\pm = \chi_c(c_\pm)}$.
Consequently, the pattern profile approaches the plateau densities $\rho_\pm$ exponentially for $x\to \pm \infty$ as
\begin{equation}\label{eq:mesa-exp-tails}
    \delta\rho_\pm^\infty = a_\pm \exp(\mp x/\ell_\pm)\, ,
\end{equation}
with $\ell_\pm = \sqrt{-D_c/\partial_c \tilde{f}|_{c_\pm}}$.

\subsubsection{Stationary pattern profile on a finite interval}
\label{sec:stat-mesa-interval}

On the finite interval ${I = [0,\Lambda/2]}$ with no-flux boundary conditions, the stationary interface profile must fulfill the boundary conditions at distances $L_\pm$ from the inflection point of the pattern, i.e., the interface region (see Fig.~\ref{fig:stat-mesa}; cf.~Ref.~\cite{Brauns.etal2021}).
These lengths $L_\pm$ of the high- and low-density plateau, respectively, are fixed by the average density $\Bar{\rho}$.
Approximating the interface as a step function (sharp-interface approximation), one has
\begin{equation}\label{eq:approx_plateau_length}
    \Bar{\rho} \, \frac{\Lambda}{2} 
    \approx 
    \rho_- L_- + \rho_+ L_+
    \, .
\end{equation}
Together with ${L_- + L_+ = \Lambda/2}$, this yields
\begin{equation}\label{eq:plateau-lengths}
    L_\pm 
    = 
    \pm \frac{\Bar{\rho}-\rho_\mp}{\rho_+-\rho_-} \, 
    \frac{\Lambda}{2}
    \, .
\end{equation}
We also define the mesa mass $M$ by
\begin{align}\label{eq:mesa-mass}
    \frac{M}{2} 
    &\equiv \int_I\mathrm{d}x\, [\rhoStat(x)-\rho_-] 
    = \frac{\Lambda}{2}(\Bar{\rho}-\rho_-) \nonumber \\
    &\approx (\rho_+-\rho_-) L_+\, .
\end{align}
where we have used Eq.~\eqref{eq:approx_plateau_length} in the last line.

Using the linearized profile equation Eq.~\eqref{eq:lin-profile-eq}, the no-flux boundary conditions imply the profiles
\begin{equation}\label{eq:mesa-cosh-tails}
    \mp [\rhoStat(x)-\rho_\pm] = \delta\rho_\pm \cosh[(x\mp L_\pm)/\ell_\pm]
\end{equation}
in the high- and low-density plateau, respectively.
The amplitudes $\delta\rho_\pm$ can be found by asymptotic matching to the exponential tails of the interface solution on the infinite domain \cite{Kolokolnikov.etal2006,Weyer.etal2023}, which yields
\begin{equation}
    \delta\rho_\pm = 2 a_\pm \exp(-L_\pm/\ell_\pm)\, .
\end{equation}
From Eq.~\eqref{eq:c-expression}, the tail amplitudes $\delta c_\pm$ of the chemoattractant profile follow as 
\begin{equation}
    \delta c_\pm = \delta\rho_\pm D_\rho/(T \rho_- \chi_c^- \chi_\rho^-)\equiv 2 a^c_\pm \exp(-L_\pm/\ell_\pm) \, .
\end{equation}

At the finite distances $L_\pm$ from the interface, the plateau densities $\rho_\pm$ are not approached exactly but one has (see Fig.~\ref{fig:stat-mesa})
\begin{equation}
    \check{\rho} = \rho_-+\delta\rho_-\, , \qquad \hat{\rho} = \rho_+ - \delta\rho_+\, ,
\end{equation}
with analogous definitions for the chemoattractant density profile $\cStat(x)$.
Thus, the total turnover balance is changed, and the stationary mass-redistribution potential $\etaStat$ for the elementary pattern on a finite domain is determined by
\begin{equation}\label{eq:ttb-mesa}
    0 = \int_{c_-+\delta c_-}^{c_+ - \delta c_+}\mathrm{d}c\, \Tilde{f}(c,\etaStat)\, .
\end{equation}
This implies that $\etaStat$ is shifted by an amount exponentially small in the plateau lengths compared to $\etaStat^\infty$.
Linearizing Eq.~\eqref{eq:ttb-mesa} in $\etaStat-\etaStat^\infty$ and $\delta c_\pm$, one obtains 
\begin{equation}\label{eq:mesa-eta-lengthDep}
    \partial_{L_\pm} \etaStat = \pm \frac{4 (a^c_\pm)^2}{\ell_\pm} \frac{\partial_c \tilde{f}|_{c_\pm}}{\Tilde{F}_\eta} \exp(-2 L_\pm/\ell_\pm)\, ,
\end{equation}
where ${\Tilde{F}_\eta = \int_{c_-}^{c_+}\mathrm{d}c\, \partial_\eta \tilde{f}(c,\etaStat^\infty)}$. The relation Eq.~\eqref{eq:mesa-eta-lengthDep} can be understood geometrically in terms of the change in the areas enclosed between the FBS and NC [see Fig.~\ref{fig:stat-mesa}(b)] \cite{Brauns.etal2021}.

In a large system with several mesas, the plateau lengths are changed dynamically by the redistribution of the cell mass within the system.
Therefore, we now determine the dependence of the mass-redistribution potential on changes in the cell mass of the mesa.
A length change $\delta L$ in the upper or lower plateau length results from a shift of the interface by $\delta L$ and thus is due to a mass change ${\delta M = \pm 2 (\rho_+-\rho_-)\delta L}$ of the mesa [cf.\ Fig.~\ref{fig:stat-mesa}(a)]. Thus, we define
\begin{equation}
    \partial_M^\pm \etaStat 
    \equiv 
    \pm\frac{1}{2(\rho_+-\rho_-)} \,
    \partial_{L_\pm} \etaStat\, .
\end{equation}
Because on a domain of fixed length, a change in the mesa mass always changes both plateau lengths ($L_+ + L_- = \mathrm{const}.$), one then has
\begin{equation}
    \partial_M^{}\etaStat 
    = 
    \partial_M^+\etaStat + \partial_M^-\etaStat\, .
\end{equation}
Note that lateral stability of the plateau densities $c_\pm$ ensures that ${\partial_c \tilde{f}|_{c_\pm} < 0}$.
Moreover, one has $\partial_\eta \tilde{f}>0$ because ${\partial_\eta \tilde{f} = T\chi_\rho\rho (\partial_\rho f)/D_\rho\geq 0}$ holds.
Thus, we find generically for mesa patterns
\begin{equation}
    \partial_M^\pm \etaStat < 0\, .
\end{equation}
This relation, as discussed below, implies that mass is generically transported from smaller to larger pattern domains, i.e., larger aggregates deplete their surrounding low-density plateau more strongly of cells than smaller aggregates and aggregates attract each other more strongly the closer they are.
These two effects induce competition and coalescence of mesa patterns that drive an uninterrupted coarsening process until all cells have accumulated in a single aggregate.

\begin{figure*}
    \centering
    \includegraphics[]{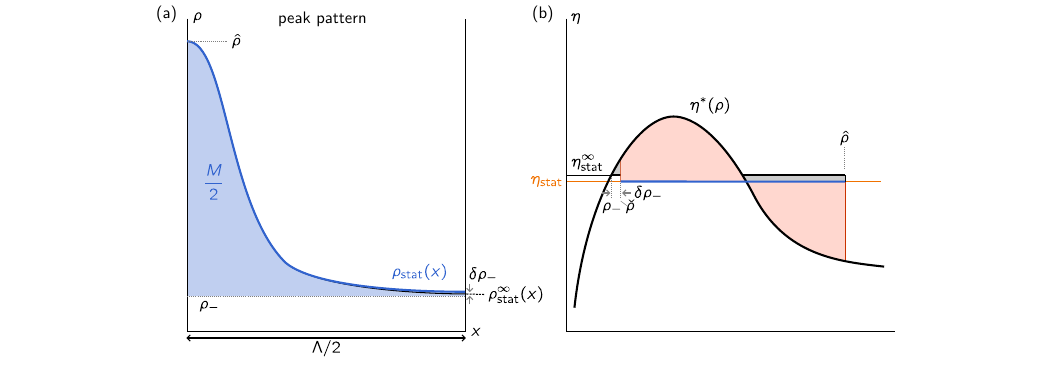}
    \caption{Construction of the stationary elementary peak pattern.
    (a) The elementary peak pattern (blue) consists of half a peak with a low-density plateau on a domain of length $\Lambda/2$ with no-flux boundary conditions.
    The peak height $\hat{\rho}$ depends on the peak mass $M$ (blue-shaded region).
    As for mesa patterns (see Fig.~\ref{fig:stat-mesa}), the pattern profile deviates in the plateau from the profile (black) on the half-infinite line $[0,\infty)$ due to the boundary condition.
    (b) Again, the pattern can be constructed in the local $(\rho,\eta)$ phase space (see Ref.~\cite{Weyer.etalinpreparation}).
    Peak patterns arise if the cell density does not saturate and does not form a high-density plateau.
    Therefore, peak patterns occur if the nullcline is $\mathsf{\Lambda}$- rather than $\mathsf{N}$-shaped and no (third) high-density intersection between the flux-balance subspace $\etaStat=\mathrm{const.}$ and the reactive nullcline $\eta^*(\rho)$ (FBS-NC intersection) exists (cf.\ Fig.~\ref{fig:stat-mesa}).
    Alternatively, peak patterns also form if the third FBS-NC intersection lies at higher densities $\rho_+\gg\hat{\rho}$ that are not reached for the given peak mass $M$.
    The value of $\etaStat$ is fixed qualitatively by the balance of the red-shaded areas between the FBS and the nullcline $\eta^*(\rho)$.
    It differs by an amount exponentially small in the plateau width from the value of the pattern on the half-infinite line due to the change of the red- compared to gray-shaded areas induced by the deviation $\delta\rho_-=\check{\rho}-\rho_-(\etaStat^\infty)$.
    The mass-redistribution potential $\etaStat$ depends, via the area balance, also on the peak height $\hat{\rho}$.
    }
    \label{fig:stat-peak}
\end{figure*}

\subsection{Peak patterns}

Peak patterns form if the cell density does not saturate in a high-density plateau (see Fig.~\ref{fig:stat-peak}).
Analogously to mesa patterns, we first analyze the stationary profile $\rhoStat^\infty(x)$ of a (half-)peak centered at ${x = 0}$ on the half-infinite domain ${I = [0,\infty)}$.
Again, the low-density plateau densities $\rho_-(\etaStat^\infty)$ and $c_-(\etaStat^\infty)$ are approached via exponential pattern tails emanating from the peak.
Importantly, for mesa patterns, a single interface on the infinite line has a unique value $\etaStat^\infty$ fixed by Eq.~\eqref{eq:ttb-mesa-infLine} that only depends on the reaction term $\tilde{f}$.
In contrast for the half-peak on the half-infinite line, the stationary mass-redistribution potential $\etaStat^\infty$ depends on the peak height $\hat{c}$ (or equivalently $\hat{\rho}$) via the total turnover balance Eq.~\eqref{eq:ttb}
\begin{equation}
    0 
    = 
    \int_{c_-(\etaStat^\infty)}^{\hat{c}}  \mathrm{d}c 
    \; \Tilde{f}(c,\etaStat^\infty)\, .
\end{equation}

As discussed below, during coarsening the peak sizes change by the transport of mass between the peaks, which is induced by gradients in $\eta$.
To determine the changes in $\eta$ with the peak mass, we first must define the mass of a peak.
Afterward, we derive the changes $\partial_M^\pm\etaStat$ of the stationary mass-redistribution potential with the peak mass.
These changes drive mass competition between peaks and determine the rates of mass competition derived in Sec.~\ref{sec:mass-comp}.
Because the profile equation on the half-infinite line with the boundary conditions ${\lim_{x\to\infty} \rhoStat^\infty = \rho_-(\etaStat^\infty)}$ and ${\lim_{x\to\infty} \partial_x \rhoStat^\infty = 0}$ has a unique solution for a fixed value of $\etaStat^\infty$ (Picard–Lindelöf theorem), the stationary profiles can be parametrized by $\etaStat^\infty$.
Thus, one can define the peak mass $M_{\infty}(\etaStat^\infty)$ by
\begin{equation}
    \frac{M_{\infty}(\etaStat^\infty)}{2} = \int_0^\infty\mathrm{d}x\, [\rhoStat^\infty(x)-\rho_-(\etaStat^\infty)]\, .
\end{equation}

On a finite domain ${I = [0,\Lambda/2]}$ which is large compared to the width of the peak (sharp-peak approximation), only the exponential profile tail of the stationary peak profile $\rhoStat(x)$ in the plateau region is modified into a $\cosh$-profile to fulfill the boundary condition [cf.~Eq.~\eqref{eq:mesa-cosh-tails} in the discussion for mesa patterns]. 
The peak profile in the narrow peak region, as does the interface profile of mesa patterns, remains unaffected (sharp-interface/-peak approximation).
The result is a change of $\etaStat$ compared to $\etaStat^\infty$ that is exponentially small in the domain length $\Lambda/2$, as we calculated explicitly for the mesa patterns [cf.\ Eq.~\eqref{eq:mesa-eta-lengthDep}].
Concomitantly, the difference between the peak mass on the infinite domain compared to the peak mass
\begin{equation}\label{eq:rho-peak-mass}
    \frac{M(\etaStat)}{2} = \int_0^{\Lambda/2}\mathrm{d}x\, [\rhoStat(x)-\rho_-(\etaStat)]
\end{equation}
on the finite domain is exponentially small in the domain length $\Lambda/2$ as well.
Analogously, we define the chemoattractant mass of the peak as
\begin{equation}\label{eq:chemoattractant-mass}
    \frac{M_c(\etaStat)}{2} = \int_0^{\Lambda/2}\mathrm{d}x\, [\cStat(x)-c_-(\etaStat)]\, .
\end{equation}
The chemoattractant mass will become relevant in the reaction-limited regime of peak competition, as derived below.

Finally, we relate the peak mass to the average cell density $\bar{\rho}$ in the domain, which is an external control parameter.
From Eq.\eqref{eq:rho-peak-mass}, we obtain
\begin{equation}
    \frac{\Lambda}{2}\bar{\rho} \approx \frac{M}{2} + \frac{\Lambda}{2} \rho_-(\etaStat)\, ,
\end{equation}
which simplifies to
\begin{equation}\label{eq:peak-mass-rhobar}
    M \approx \Lambda [\Bar{\rho}-\rho_-(\etaStat)]\, .
\end{equation}
The Supplemental Material to Ref.~\cite{Brauns.etal2021} discusses that for the stable elementary patterns undergoing coarsening, the function $\etaStat(M)$ is single-valued [as is $M(\etaStat)$] and the stationary peak profiles can be parametrized by the peak mass $M$ instead of $\etaStat$.
With this, we define $\rho_-(M) = \rho_-(\etaStat^\infty(M))$ and $\hat{\rho} = \hat{\rho}(M)$, $\hat{c} = \hat{c}(M)$.

As discussed above, the difference of the stationary mass-redistribution potential $\etaStat$ to its value $\etaStat^\infty$ for a peak with the same mass $M$ on the half-infinite domain is exponentially small in the domain length because the change is only due to the exponential pattern tails (see Fig.~\ref{fig:stat-peak}).
As for the mesa patterns [cf.~Eq.~\eqref{eq:mesa-eta-lengthDep}], one defines for stationary peak patterns with fixed $\hat{\rho}$ (within the sharp-peak approximation)
\begin{subequations}\label{eq:peak-eta-m}
\begin{align}
    \partial_M^-\etaStat &\equiv -\frac{1}{2[\hat{\rho}(M)-\rho_-(M)]} \partial_{L_-}\etaStat \\
    &\equiv -\frac{1}{2(\hat{\rho}-\rho_-)} \partial_{\Lambda/2}\etaStat \\
    &= \frac{1}{\hat{\rho}-\rho_-} \frac{2 (a_-^c)^2}{\ell_-} \frac{\partial_c \tilde{f}|_{c_-}}{\Tilde{F}_\eta} \exp(- \Lambda/\ell_-)\, ,
\end{align}
\end{subequations}
with 
\begin{equation}
\label{eq:Ftilde}
    \Tilde{F}_\eta \equiv \int_{c_-}^{\hat{c}(M)}\mathrm{d}c\, \partial_\eta \tilde{f}(c,\etaStat^\infty(M))
    \, .
\end{equation}
As for mesa patterns, one has ${\partial_M^-\etaStat < 0}$. 

Changing only the domain length, we keep the heights $\hat{\rho}(M)$ and $\hat{c}(M)$ fixed when calculating $\partial_M^-\etaStat$ in Eq.~\eqref{eq:peak-eta-m}.
The heights $\hat{\rho}(M)$ and $\hat{c}(M)$ change if the peak mass is changed (beyond the change of the exponential tail in the case of a changing domain length).
This occurs if the average cell density $\Bar{\rho}$ is changed in a finite system.
As for the change of the exponential tails in the plateaus, these height changes alter the total turnover balance and thereby the stationary mass-redistribution potential.
Ref.~\cite{Weyer.etal2023} gives an approximate scaling argument for large peak masses that yields a power-law dependence ${\etaStat\approx \etaStat^\infty\sim M^{-\alpha}}$ deriving from the scaling of the total turnover balance at large densities.
Within the sharp-peak approximation, the change in the length of the low-density plateau on a large, finite domain is negligible when the peak mass is changed, and we define
\begin{equation}
    \partial_M^+\etaStat \equiv \partial_{M_{\infty}}^{} \etaStat^\infty \approx \partial_M^{} \etaStat\, .
\end{equation}
Instead of the length of the high-density plateau, here the derivative $\partial_M^+$ signifies a change of the ``size'' of the peak.
To argue that coarsening is generically uninterrupted, an argument is given in the Supplemental Material of Ref.~\cite{Brauns.etal2021} that $\partial_M\etaStat$ remains negative along a branch of stable elementary peak patterns.
The argument also applies to the peak patterns in the Keller--Segel systems discussed here.
In Ref.~\cite{Weyer.etalinpreparation}, we assume that the peak height growth with the peak mass, which is sufficient to ensure ${\partial_M\etaStat < 0}$.

In the mKS system, the stationary profiles in the limit of large peak masses $M$ have been determined approximately in Ref.~\cite{Kang.etal2007} using asymptotic matching.
In our notation, the cell-density profile in the inner peak region [${x = \mathcal{O}(1/M)}$] reads
\begin{equation}\label{eq:mKS-stat-rho}
    \rhoStat^\infty(x) \approx \frac{M^2 T}{8 D_\rho D_c} \operatorname{sech}^2\left(\frac{M T}{4 D_\rho D_c}x\right),
\end{equation}
and the chemoattractant density is
\begin{equation}\label{eq:mKS-stat-c-inner}
    \cStat^\infty(x) \approx - \frac{D_\rho}{T} \log\left[4 \cosh^2\left(\frac{M T}{4 D_\rho D_c}x\right)\right] + \frac{M}{2 \sqrt{D_c}}\, .
\end{equation}
In the outer region [$x\gg \mathcal{O}(1/M)$], the chemoattractant density is given by
\begin{equation}\label{eq:mKS-stat-c-outer}
    \cStat^\infty(x) \approx \frac{M}{2 \sqrt{D_c}} \exp\left(-x/\sqrt{D_c}\right).
\end{equation}

Determining the stationary mass-redistribution potential $\etaStat^\infty$ in the inner peak region using its definition Eq.~\eqref{eq:mKS-eta} and Eqs.~\eqref{eq:mKS-stat-rho},~\eqref{eq:mKS-stat-c-inner}, one obtains
\begin{equation}\label{eq:mKS-stat-eta}
    \etaStat^\infty \approx \frac{D_\rho}{T} \log \left(\frac{M^2 T}{2 D_\rho D_c}\right) - \frac{M}{2 \sqrt{D_c}}\, ,
\end{equation}
that is, asymptotically for large peak masses one has ${\etaStat^\infty \sim - M}$.

Taken together, we classify the patterns of Keller--Segel models into mesa (see Fig.~\ref{fig:stat-mesa}) and peak patterns (see Fig.~\ref{fig:stat-peak}).
The difference is that in peak patterns the cell density does not saturate and does not form a high-density plateau.
The mass of mesa and peak patterns, that is, their width and peak size, is related to the average cell density $\bar{\rho}$, which is set by the initial condition for a single half-mesa or half-peak.
However, on a large domain with several peaks and mesas, the peak and mesa masses can change by the mass transport between different peaks and mesas.
This occurs during coarsening and is the basis of the mass-competition instability discussed in the next section.
For this instability, it will be crucial that patterns of different masses have different stationary values $\etaStat$ of the mass-redistribution potential [cf.\ Figs.~\ref{fig:stat-mesa}(b),~\ref{fig:stat-peak}(b)].
In both mesa and peak patterns, it changes by an exponentially small amount with the plateau widths [cf.\ Eqs.~\eqref{eq:mesa-eta-lengthDep},~\eqref{eq:peak-eta-m}].
Relating the changes in the high- or low-density plateau lengths to changes in the cell mass, one finds ${\partial_M^\pm\etaStat < 0}$.
For peak patterns, no high-density plateau exists and $\partial_M^+\etaStat$ is defined as the change of $\etaStat$ with the peak mass, and the change ${\partial_M^+\etaStat < 0}$ is negative as well if the peak height grows with the peak mass.

\section{Mass-competition instability}
\label{sec:mass-comp}

\begin{figure}
\centering
\includegraphics[width=\columnwidth]{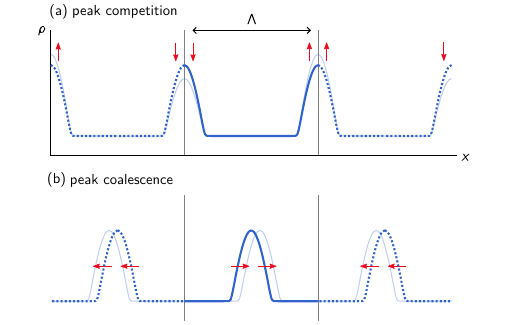}
\caption{Coarsening scenarios of a peak pattern.
(a) \textit{Peak competition}: Peak patterns (dark-blue profile) can undergo coarsening by the transport of mass from smaller to larger peaks, resulting in the collapse of the smaller peaks (red arrows and light-blue profile).
This competition for mass between peaks renders stationary periodic patterns with equally large peaks unstable to small disturbances in the peak masses.
Due to the symmetry of the perturbation mode, peak competition can be analyzed by the interaction of two half peaks on a domain of length $\Lambda$ with no-flux boundary conditions (gray lines).
(b)  \textit{Peak coalescence}: 
Peak patterns can also undergo coarsening due to the coalescence of peaks (red arrows and light-blue profile).
This peak coalescence renders stationary periodic pattern unstable to small variations in the peak distances.
It can be analyzed by examining the movement of a single peak on a domain of length $\Lambda$ with no-flux boundary conditions (gray lines).
The coalescence with the neighboring peak of the periodic pattern then corresponds to the coalescence with the boundary.}
\label{fig:coarsening-scenarios}
\end{figure}

In this section, we derive the rates of the mass-competition instability for peak coalescence and peak competition [Fig.~\ref{fig:coarsening-scenarios}] using \textit{singular perturbation theory}.
The calculations directly generalize to mesa patterns as shown explicitly for two-component mass-conserving reaction--diffusion systems in Ref.~\cite{Weyer.etal2023}.
The resulting expressions for the rates are provided at the end of this section.

How is the mass-competition instability related to the coarsening of patterns with several peaks or mesas?
Coarsening can proceed by two processes.
First, larger pattern domains (peaks or mesas) grow while smaller domains collapse and vanish.
Second, pattern domains closer to each other than to their other neighbors approach even further until they coalesce.
Both processes lead to a reduction in the number of pattern domains, giving rise to the coarsening process of the pattern.

In each of these two scenarios, the coarsening process can be understood as an instability of a stationary periodic pattern with equally-sized peaks or mesas.
In the first scenario, which we will refer to as \textit{peak} or \textit{mesa competition}, small perturbations to the peak and mesa masses grow and destabilize the periodic pattern [see Fig.~\ref{fig:coarsening-scenarios}(a)].
This process is similar to Ostwald ripening of droplets in phase-separating mixtures, which induces the growth of larger and shrinking of smaller droplets \cite{Lifshitz.Slyozov1961,Wagner1961}.
In the second scenario, referred to as \textit{peak} or \textit{mesa coalescence}, the stationary periodic pattern become destabilized by the amplification of small deviations in the distances between peaks [see Fig.~\ref{fig:coarsening-scenarios}(b)].

As our mathematical analysis will explicitly show, both coarsening processes are driven by self-amplifying mass transport induced by the changes in peak and mesa masses and plateau lengths.
Therefore, we refer to the instability as the \emph{mass-competition instability} (cf.\ Ref.~\cite{Brauns.etal2021,Weyer.etal2023} for the analogous instability in mass-conserving reaction--diffusion systems).
Because mass transport between peaks or mesas, as well as coalescence, is expected to be fastest on the shortest distances, we consider the competition between neighboring peaks [cf.\ Fig.~\ref{fig:coarsening-scenarios}].
On the basis of the calculated growth rates of the mass-competition instability, we then determine the coarsening law, that is, the temporal evolution of the average pattern wavelength, through a scaling argument~\cite{Langer1971,Glasner.Witelski2003,Weyer.etal2023,Weyer.etalinpreparation}.

In Ref.~\cite{Weyer.etalinpreparation}, we constructed the growth rates of the mass-competition instability within a QSS approximation for the mass-redistribution potential at the individual peaks.
As in two-component mass-conserving reaction--diffusion systems \cite{Weyer.etal2023}, we show here that the QSS approximation holds in the diffusion-limited regime.
The growth rates also show a reaction-limited regime, where the local reactive dynamics of the chemoattractant, rather than mass transport—either between the peaks during peak competition or through the peaks during peak coalescence—becomes the rate-limiting factor.
Importantly, in both the diffusion- and reaction-limited regimes, the instability criterion turns out to be the same.
In each case, the mass-competition instability occurs when the stationary mass-redistribution potential decreases with the (associated) domain mass, ${\partial_M^\pm \etaStat < 0}$.

\subsection{Mass-competition rates from singular perturbation theory}
\begin{figure*}[htb]
    \centering
    \includegraphics[]{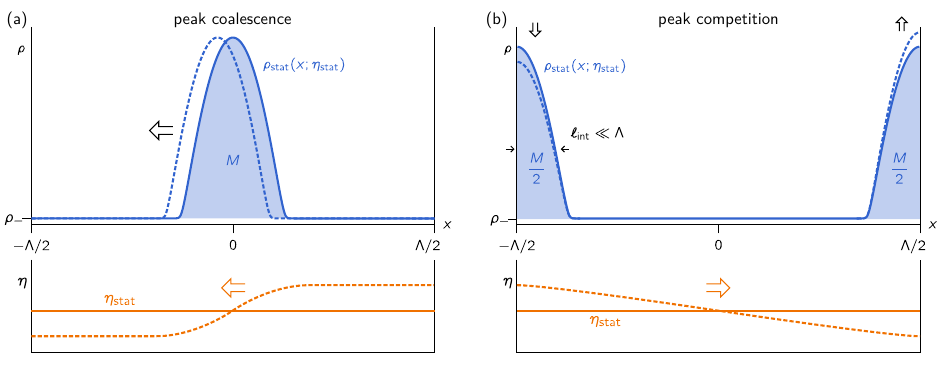}
    \caption{
    Mass-competition scenarios.
    (a) Coalescence of neighboring peaks of a periodic pattern (solid blue profile) corresponds to the coalescence of a single peak with a reflective boundary (black arrow and dashed profile).
    We calculate the growth rate of a small shift of the peak position away from the center ${x = 0}$ of the domain ${I = [-\Lambda/2,\Lambda/2]}$ with reflective boundary conditions.
    The coalescence scenario of the mass-competition instability can be understood as the competition for (negative) mass of the two low-density plateaus mediated by the gradient $\eta(x)$ of the mass-redistribution potential induced within the peak (dashed orange profile in the lower panel) \cite{Brauns.etal2021,Weyer.etal2023}.
    The stationary mass-redistribution potential $\etaStat$ is constant.
    The plateau has approximately the density $\rho_-$.
    (b) Alternatively, neighboring peaks interact by competing for mass (black arrows).
    We calculate the growth rate of a small mass difference between the two half peaks with mass $M/2$ at the left and right reflective boundary of the domain ${I = [-\Lambda/2,\Lambda/2]}$ (blue profile).
    Mass transport between the peaks is due to the approximately linear gradient $\eta(x)$ (dashed orange profile in the lower panel).
    To calculate both growth rates, we assume large peak masses $M$ (blue-shaded area) and a separation between the peaks $\Lambda$ (pattern wavelength) that is large compared to the interface (half-peak) width $\ell_\mathrm{int}$.
    This separation corresponds to the distance between the peak and its image mirrored at the reflective boundary in the coalescence scenario.
    }
    \label{fig:massComp-scenarios}
\end{figure*}

As the basic mechanism driving the coarsening of a pattern with multiple domains, we consider the interaction of two (equal) elementary patterns of width $\Lambda/2$ (half-peak or half-mesa) on a domain $[-\Lambda/2,\Lambda/2]$ with no-flux boundaries [Fig.~\ref{fig:massComp-scenarios}].
Using these elementary patterns, one can construct two different representations (``unit cells'') of a periodic pattern with wavelength $\Lambda$ on a larger domain: either by combining them to form a single centered peak [Fig.~\ref{fig:massComp-scenarios}(a)] or by arranging them as two half-peaks [Fig.~\ref{fig:massComp-scenarios}(b)] positioned at the domain boundaries.
These representations are used in the following to analyze the stability of stationary periodic patterns with respect to two distinct mass-competition scenarios. It is shown that this leads either to coalescence of peaks (with their mirror image or a neighbor on a larger domain) [Fig.~\ref{fig:massComp-scenarios}(a)] or to peak competition [Fig.~\ref{fig:massComp-scenarios}(b)].

To determine the time evolution of small perturbations to a symmetric stationary pattern $[\rhoStat(x),\etaStat]$, we set ${\rho(x,t) = \rhoStat(x) + \delta\rho(x,t)}$ and ${\eta(x,t) = \etaStat + \delta\eta(x,t)}$, and linearize the Keller--Segel dynamics, Eqs.~\eqref{eq:KS}, around the stationary pattern. This yields [using $\rho$ and $\eta$ as independent variables, i.e., Eqs.~\eqref{eq:cont-eq},~\eqref{eq:KS-eta}]
\begin{equation}\label{eq:linearized-dynamics}
    \partial_t \begin{pmatrix}
    \delta\rho(x,t) \\
    \delta\eta(x,t)
    \end{pmatrix}
    = 
    \mathbf{L}(\rho_\mathrm{stat}(x), \eta_\mathrm{stat})
    \begin{pmatrix}
     \delta\rho(x,t) \\
     \delta\eta(x,t)
    \end{pmatrix}.
\end{equation}
The operator $\mathbf{L}$ on the right-hand side contains the linearized diffusion and chemotaxis terms.
Moreover, it contains the linearized reaction term of the chemoattractant.
The explicit form of the operator is given below.

The goal is to calculate the growth rate $\sigma$ of the mass-competition instability.
It is given by the eigenmode $[\delta \rho(x,t),\delta\eta(x,t)] = \operatorname{e}^{\sigma t}[\delta \rho(x),\delta\eta(x)]$ that describes the coalescence or competition of the peaks.
For convenience, we will denote $[\delta \rho(x),\delta\eta(x)]$ by $[\delta \rho,\delta\eta]$, omitting the arguments.
Inserting this ansatz into Eq.~\eqref{eq:linearized-dynamics}, we have to solve the eigenvalue problem
\begin{equation} 
\label{eq:comp-eigenvalue-problem}
\sigma \begin{pmatrix}
    \delta\rho \\
    \delta\eta
\end{pmatrix}
 = 
 \mathbf{L}(\rho_\mathrm{stat}(x), \eta_\mathrm{stat})
 \begin{pmatrix}
     \delta\rho \\
     \delta\eta
 \end{pmatrix} .
\end{equation}
To distinguish the growth rate of peak competition [Fig.~\ref{fig:coarsening-scenarios}(a)] from peak coalescence [competition between the low-density plateaus; Fig.~\ref{fig:coarsening-scenarios}(b)], we will denote their growth rates by $\sigma^\pm$, respectively.
As the following expressions in this subsection do not depend on the type of mass-competition processes, we will denote the growth rate, which may either be the competition or coalescence rate, by $\sigma$.

Unlike the linear stability analysis for the homogeneous steady state (c.f.\ Sec.~\ref{sec:lsa}), the evolution operator of the linearized dynamics $\mathbf{L}$ is explicitly space-dependent.
This spatial dependence arises because the reaction rates and the linearized chemotaxis term in Eqs.~\eqref{eq:cont-eq},~\eqref{eq:KS-eta} depend on the stationary cell density $\rhoStat(x)$, which is space-dependent.
As a result, the eigenmodes are not simply Fourier modes. 
Instead, the spatial profile of the eigenmodes must be determined alongside the growth rate $\sigma$.
To address this difficulty, we will exploit the fact that mass competition is slow compared to the relaxation of the pattern profile onto the stationary profile and determine $\sigma$ perturbatively by approximating the eigenmodes by (approximate) zero modes of the linear dynamics in Secs.~\ref{sec:peak-coal},~\ref{sec:peak-comp}.

Finally, to simplify the analysis of the eigenvalue problem Eq.~\eqref{eq:comp-eigenvalue-problem}, we will use the perturbation of the chemoattractant density $\delta c$ that can be determined from $\delta\rho$ and $\delta\eta$ by linearizing Eq.~\eqref{eq:c-expression}.
This yields
\begin{equation}\label{eq:massComp-delta-c}
    \delta c 
    = 
    \frac{D_\rho}{T\chi_c\chi_\rho \rhoStat} \, \delta \rho 
    - 
    \frac{1}{\chi_c} \, \delta\eta
    \, .
\end{equation}
Here and in the following, all quantities $\chi_c$, $\chi_\rho$, and $f_{\rho,\eta,c}$ are assumed to be evaluated along the stationary pattern $[\rhoStat(x),\etaStat]$ if not noted differently.
With Eq.~\eqref{eq:massComp-delta-c}, the eigenvalue problem Eq.~\eqref{eq:comp-eigenvalue-problem} can be written compactly as\footnote{
This is most easily seen by writing the generalized Keller--Segel system Eq.~\eqref{eq:KS} as
\begin{align}
    \partial_t \rho &= \partial_x[T\chi_\rho \rho \partial_x \eta],\nonumber\\
    \partial_t c &= D_c \partial_x^2 c + \tilde{f}(c,\eta),\nonumber
\end{align}
using the definition Eq.~\eqref{eq:f-expressions}.
Linearizing this equation in $\rho$, $c$, and $\eta$, one obtains Eq.~\eqref{eq:massComp-eigenvalue-problem}.
}
\begin{subequations}\label{eq:massComp-eigenvalue-problem}
\begin{align}
    \sigma \, \delta \rho 
    &= \partial_x \big[ T\chi_\rho \rhoStat \partial_x \delta \eta \big] \, ,
    \label{eq:massComp-lin-cont-eq}\\
    \sigma \, \delta c 
    &= D_c \partial_x^2 \delta c + (\partial_c \tilde{f}) \, \delta c + (\partial_\eta \tilde{f}) \, \delta \eta \, .
    \label{eq:massComp-lin-c-eq-1}
\end{align}
\end{subequations}
In addition, we define the linear operator
\begin{equation}
    \mathcal{L} \equiv - D_c \partial_x^2 - \partial_c \tilde{f}\, .
\end{equation}
Using this definition, Eq.~\eqref{eq:massComp-lin-c-eq-1} reads
\begin{equation}\label{eq:massComp-lin-c-eq}
    (\partial_\eta \tilde{f}) \, \delta \eta - \sigma \, \delta c =  \mathcal{L} \, \delta c\, .
\end{equation}

\subsubsection{Assumptions and approximations for solving the eigenvalue problem}
\label{sec:assumptions}
Because the evolution operator of the linear dynamics is explicitly space-dependent, the eigenmodes cannot be determined in general.
To solve the eigenvalue problem in Eqs.~\eqref{eq:massComp-eigenvalue-problem} and determine the growth rate $\sigma$, one must approximate the spatial profile of the eigenmode.
To this end, we use singular perturbation theory and perform an analysis similar as in Refs.~\cite{Kang.etal2007,Brauns.etal2021,Weyer.etal2023}.
In this subsection, we detail the approximations we will use in the singular perturbation analysis performed below in Secs.~\ref{sec:peak-coal},~\ref{sec:peak-comp}.

We assume that the peak mass is large and the peaks are narrow compared to the domain length $\Lambda$ (sharp-interface approximation; ``coarsening limit'').
Similarly, for mesa patterns we assume that the interface width is narrow compared to the domain lengths.
In these limits, the stationary profiles and the eigenmode profiles can be constructed using asymptotic matching of the peak or interface profile and the profile of the plateaus (cf.\ Sec.~\ref{sec:stat-patterns}).
Thus, on the scale of the domain length, one analyzes the limit of the peak becoming infinitely sharp (the interface becoming a step function).
This limit is a singular limit because the pattern profile only converges pointwise to these limit profiles.
Therefore, the analysis within the sharp-interface (sharp-peak) approximation is called singular perturbation theory.

In addition, we assume that the relaxation to the stationary, elementary peak profile (i.e., for each single peak) is fast compared to the rate of mass competition.
This separation of timescales has been discussed explicitly for coarsening in two-component mass-conserving reaction--diffusion systems in Ref.~\cite{Weyer.etal2023}, and we recapitulate the discussion for the Keller--Segel systems in Appendix~\ref{sec:elementary-stab}.
This analysis shows that fast relaxation of elementary patterns compared to the timescale of mass competition requires
\begin{equation}\label{eq:plateau-negligible}
    |\partial_{\etaStat} M|
    \gg
    \Lambda \, \partial_{\etaStat} \rho_-
    \, ,
\end{equation}
that is, the change in the plateau mass must be negligible compared to the change in the peak mass.
With the condition Eq.~\eqref{eq:plateau-negligible}, one can neglect any changes in the low-density plateau and only consider the redistribution of mass between the peaks.
This is similar to the observation that one can neglect the changing supersaturation when determining the droplet mass during Ostwald ripening \cite{Lifshitz.Slyozov1961,Wagner1961}.
The condition Eq.~\eqref{eq:plateau-negligible} is generically fulfilled if the peak mass of (stable) stationary peaks is increased sufficiently, i.e., ${M \to \infty}$, because the lateral stability of the low-density plateau requires ${\partial_{\etaStat} \rho_- > 0}$ (cf.\ Sec.~\ref{sec:lsa}) which results in ${\partial_M \rho_- < 0}$ for peaks undergoing coarsening, that is, if the peaks fulfill ${\partial_M^\pm\etaStat < 0}$ (chain rule).
Moreover, the cell density is bounded from below by ${\rho\geq 0}$.
Together, ${\partial_M \rho_- < 0}$ and ${\rho\geq 0}$ imply that one has ${|\partial_M \rho_-| \to 0}$ as ${M \to \infty}$. 

For the singular perturbation analysis, we will not only need that changes in the plateau of the cell density are negligible, but also that we can neglect changes in the plateau of the chemoattractant density compared to the chemoattractant mass $M_c$ of the peak.
This approximation is possible under the same conditions as for the total cell mass $M$ if the chemoattractant mass $M_c$ increases with the peak or mesa size.
Because the chemoattractant is produced by the cells, we expect that the chemoattractant density increases with the cell density.
We therefore assume that the chemoattractant mass $M_c$ grows with the cell mass $M$, that is,\footnote{
Since the stationary profiles $\rhoStat$ and $\cStat$ are related through the constant mass-re\-dis\-tri\-bu\-tion potential $\etaStat$, one finds [cf.\ Eq.~\eqref{eq:c-expression}]
\begin{equation}\label{eq:rhostat-cstat-rel}
    \partial_{\etaStat} \cStat = [D_\rho(\partial_{\etaStat}\rhoStat)/(T\chi_\rho\rhoStat) - 1]/\chi_c \, .
\end{equation}
Because it cannot be ensured that the integral of the right-hand side of Eq.~\eqref{eq:rhostat-cstat-rel} over an elementary pattern is negative, we cannot show in general that the chemoattractant mass grows with the cell mass.}
\begin{equation}\label{eq:assumption-chemoattrMass-increase}
    \partial_M^{} M_c > \delta > 0
    \, ,
\end{equation}
where $\delta$ is a (small) constant independent of $M$. 
This assumption is generically fulfilled for linear chemoattractant production and degradation ${f(\rho,c)=k_\mathrm{p}\rho-k_\mathrm{d} c}$, which follows from integrating the chemoattractant-profile equation Eq.~\eqref{eq:profile-eq-c} over an elementary pattern.
Moreover, as discussed below, it generically holds for mesa patterns.
Assuming that the derivative $\partial_M M_c$ remains bounded away from zero, the same argument as for $\rho_-$ ensures that
\begin{equation}\label{eq:plateau-negligible-c}
    |\partial_{\etaStat} M_c|\gg\Lambda \partial_{\etaStat} c_-
\end{equation}
holds as the mass $M$ of (stable) stationary elementary patterns is increased sufficiently.

For mesa-forming systems, Eqs.~\eqref{eq:plateau-negligible}--\eqref{eq:plateau-negligible-c} hold without additional assumptions if the interfaces are sufficiently far apart (sharp-interface approximation). 
On the one hand, ${\partial_{\etaStat}\rho_-\to\partial_{\etaStat}\rho_-|_{\etaStat^\infty}}$ approaches a constant because the stationary mass-redistribution potential $\etaStat$ approaches its value $\etaStat^\infty$ of a single interface on the infinite line in the sharp-interface limit [cf.\ Eq.~\eqref{eq:mesa-eta-lengthDep}].
On the other hand, both the cell and chemoattractant masses, $M$ and $M_c$, are proportional to the length of the high-density plateau $L_+$ [see Eq.~\eqref{eq:mesa-mass}].
With the dependence of the stationary mass-redistribution potential $\etaStat$ on the plateau lengths, Eq.~\eqref{eq:mesa-eta-lengthDep}, one thus has that $|\partial_{\etaStat} M|, |\partial_{\etaStat} M_c|$ increase exponentially.
Thus, the left-hand sides of Eqs.~\eqref{eq:plateau-negligible},~\eqref{eq:plateau-negligible-c} grow exponentially with the distance between the interfaces while the right-hand side approaches a constant.
Consequently, the estimates Eqs.~\eqref{eq:plateau-negligible},~\eqref{eq:plateau-negligible-c} hold in the sharp-interface limit.

Using the discussed assumptions, we approximate the peak-coalescence mode and determine its growth rate in the following section Sec.~\ref{sec:peak-coal}.
We then analyze the peak-competition scenario in Sec.~\ref{sec:peak-comp}.

\subsubsection{Peak coalescence}
\label{sec:peak-coal}

To study peak coalescence, we consider a stationary peak centered at ${x = 0}$ on the interval ${I = [-\Lambda/2,\Lambda/2]}$ with reflective boundary conditions [Fig.~\ref{fig:massComp-scenarios}(a)].
The coalescence mode of the mass-competition instability corresponds to the translation of the peak away from the center towards one of the boundaries.
Because the stationary profile is parity symmetric (${x\to -x}$), the eigenmodes are either symmetric or antisymmetric.
To describe the translation of the peak, the coalescence mode must be antisymmetric.

\medskip

\paragraph{Approximation of the coalescence mode.\;---}
Assuming the peak is narrow compared to the domain length $\Lambda$ (sharp-peak approximation) and that relaxation to the stationary peak profile is rapid relative to the rate of mass competition (cf.\ Sec.~\ref{sec:assumptions}), we approximate the peak coalescence mode by shifting the stationary profile by a small amount $A$ such that 
\begin{equation}
    \rho (x) 
    \approx \rhoStat(x + A) 
    \approx \rhoStat(x) + A\, \partial_x \rhoStat (x)
    \, .   
\end{equation}
Thus, the peak profile of the coalescence mode is approximated by
\begin{equation}\label{eq:deltaRho-coalescence-approx}
    \delta \rho(x) 
    \approx 
    A \, \partial_x\rhoStat(x)
    \, ,
\end{equation}
where the mode amplitude is denoted by $A$.
Importantly, since the translation mode $\partial_x\rhoStat(x)$ shifts the profile as a whole it does not fulfill the no-flux boundary conditions and, therefore, must be modified close to the domain boundaries.
In more detail, in the plateau regions close to the domain boundaries, we have to consider that a shift of the peak position results in a shortening or elongation of the plateau (because the boundary remains fixed and does not shift along with the peak).
Again using that the relaxation onto the stationary peak profile is fast compared to the rate of mass-competition (cf.\ Sec.~\ref{sec:assumptions}), we approximate the coalescence mode close to the boundaries by the change of the stationary profile when changing the plateau length $L_-$.
Below, we show how to calculate this change in the plateau region.
We use the analogous approximations for the chemoattractant mode profile $\delta c$.

During peak coalescence, mass must be redistributed from one side of the peak to the other for the peak to shift its position.
This redistribution is driven by gradients in $\delta\eta$ within the peak [cf.\ Eq.~\eqref{eq:cont-eq} and Fig.~\ref{fig:massComp-scenarios}].
The profile of the mass-redistribution potential $\delta \eta$ follows from integrating the linearized continuity equation Eq.~\eqref{eq:massComp-lin-cont-eq} twice after inserting the mode approximation Eq.~\eqref{eq:deltaRho-coalescence-approx}.  
This yields
\begin{equation}\label{eq:coalescence-eta-approx}
    \delta\eta(x) 
    \approx 
    A\, \frac{\sigma}{T} \int_0^x\mathrm{d}y \, 
    \frac{\rhoStat(y)-\rho_-}{\chi_\rho(\rhoStat(y)) \, \rhoStat(y)}\, .
\end{equation}
Here, the plateau density $\rho_-$ is the integration constant of the first integration, which is chosen such that the mode profile $\delta\eta(x)$ approximately fulfills the no-flux boundary conditions, that is, ${\partial_x\delta\eta|_{\Lambda/2}\propto \rhoStat(\Lambda/2)-\rho_-\approx 0}$ (sharp-peak approximation).
By inserting the ansatz for $\delta \rho$, Eq.~\eqref{eq:deltaRho-coalescence-approx} , we neglected the corrections to this approximation close to the domain boundaries, which are exponentially small in the domain length $\Lambda$ (cf.\ Sec.~\ref{sec:stat-mesa-interval}).

Taken together, the approximation for the coalescence mode Eq.~\eqref{eq:deltaRho-coalescence-approx}, the equivalent equation for $\delta c$, the approximations close to the boundary discussed below, and Eq.~\eqref{eq:coalescence-eta-approx} constitute our approximation of the coalescence eigenmode.

\medskip

\paragraph{Determining the coalescence growth rate.\;---}
We will use this ansatz in the following to calculate the growth rate $\sigma^-$ [cf.\ Fig.~\ref{fig:coarsening-scenarios}(b)].
For simplicity of notation, we will drop the superscript ${``-"}$ for the rest of the derivation and only introduce it in the final result again.
Having used the continuity equation, Eq.~\eqref{eq:massComp-lin-cont-eq}, to determine the mode profile for the mass redistribution potential above, we need a second condition that relates $\delta\rho$, $\delta c$, and $\delta\eta$ to obtain an equation for the growth rate $\sigma$.
This is given by the second equation of the linearized dynamics, that is, the linearized chemoattractant equation Eq.~\eqref{eq:massComp-lin-c-eq}.
By multiplying this equation with $\partial_x \cStat$ and integrating the resulting expression over the domain half $[0,\Lambda/2]$, we project the chemoattractant dynamics onto the translation mode.
Using the approximations from Eqs.~\eqref{eq:deltaRho-coalescence-approx},~\eqref{eq:coalescence-eta-approx} for the coalescence mode $(\delta \rho, \delta \eta)$,
this projection yields
\begin{align} \label{eq:coal-deriv-1}
    \int_0^{\Lambda / 2}\mathrm{d} x\, 
    &(\partial_x \cStat) (\partial_\eta \tilde{f}) \, \delta \eta
    - \sigma \int_0^{\Lambda / 2}\mathrm{d}x\, (\partial_x \cStat) \, \delta c \nonumber\\
    &=
    \int_0^{\Lambda / 2}\mathrm{d}x\, (\partial_x \cStat) \, \mathcal{L} \, \delta c\,  .
\end{align}
Except for the exponential profile tails in the plateau regions, $\partial_x\cStat(x)$ is localized to the center of the domain $[-\Lambda/2,\Lambda/2]$.
Thus, on the left-hand side, one can insert the approximation Eq.~\eqref{eq:coalescence-eta-approx} for $\delta\eta$ and use ${\delta c \approx A \, \partial_x \cStat}$.
Moreover, calculating the derivative of the stationary profile equation Eq.~\eqref{eq:profile-eq-c} with respect to $x$ implies
\begin{equation}
    0 = \mathcal{L}\, \partial_x\cStat\,.
\end{equation}
This equation holds since the system is (up to the boundaries) translationally invariant.
Therefore, the shifted profile ${\cStat(x+A)\approx\cStat(x) + A \, \partial_x\cStat(x)}$ is also a solution to the stationary profile equation (only it does not fulfill the boundary conditions).

Consequently, integrating by parts twice on the right-hand side of Eq.~\eqref{eq:coal-deriv-1} yields
\begin{align}\label{eq:coales-sigma-eq-1}
    \int_0^{\Lambda / 2}\mathrm{d}x\, 
    &(\partial_x \cStat)  (\partial_\eta \tilde{f}) \, \delta\eta - \sigma A \int_0^{\Lambda / 2}\mathrm{d} x\,  (\partial_x \cStat)^2  
    \nonumber\\
    &\approx
    \left .-D_c (\partial^2_x \cStat) \, \delta c \right|_{\Lambda / 2}\, ,
\end{align}
where we have not inserted the approximation Eq.~\eqref{eq:coalescence-eta-approx} for $\delta\eta$ to ease the notation. 
The other three boundary terms arising from the two integrations by parts are not shown on the right-hand side of Eq.~\eqref{eq:coales-sigma-eq-1} since these vanish because one has ${\partial_x\cStat|_{0,\Lambda/2} = 0}$ and ${\delta c(0) = 0}$ due to the antisymmetry of the coalescence mode.

Equation~\eqref{eq:coales-sigma-eq-1}, together with $\delta\eta$ approximated by Eq.~\eqref{eq:coalescence-eta-approx}, gives an expression for the growth rate $\sigma$ that is entirely formulated in terms of the stationary pattern profile, provided we insert a suitable approximation for $\delta c|_{\Lambda/2}$.
Specifically, because this quantity is evaluated at the domain boundary, the approximation $\delta c \approx\partial_x\cStat$ is not suitable since it does not satisfy the boundary conditions at the domain ends.
Instead, we must consider that in the plateau region close to the boundary, shifting the peak results in a change in the length of the pattern plateau.
To account for this, we construct this change in the following by starting with the stationary plateau profile.
In Sec.~\ref{sec:stat-patterns}, we derived the profile of the exponential pattern tail in the plateau regions of mesa patterns [cf.\ Eq.~\eqref{eq:mesa-cosh-tails}] by asymptotic matching. 
By the same matching argument, one analogously finds for the right plateau of the peak pattern considered
\begin{align} \label{eq:coal-tail-approx}
    \cStat(x) 
    &\approx c_- 
    + 2 \, a_- 
    \exp 
    \bigg[
    -\frac{\Lambda}{2 \ell_-}
    \bigg]
    \cosh  
    \bigg[
    \frac{{\Lambda}/{2} - x}{\ell_-}  \bigg]\nonumber
    \\
    &\equiv c_- + \delta c_- \, 
    \cosh  
    \bigg[
    \frac{{\Lambda}/{2} - x}{\ell_-} 
    \bigg]
    \, ,
\end{align}
where ${\delta c_- = \check{c}-c_-}$ denotes the amplitude of the profile tail at the domain boundary.
As discussed in Sec.~\ref{sec:stat-mesa-interval} for mesa patterns, here the amplitude $a_-$ of the tail depends on the profile of the peak on the half-infinite line.

During the coalescence process the peak is shifted from ${x = 0}$ to ${x = -A}$ [cf.\ Eq.~\eqref{eq:deltaRho-coalescence-approx}].
Approximating the resulting profile by the stationary profile, it must be composed of a stationary elementary half-peak patterns on the domain $[\Lambda/2,-A]$ and a (reflected) elementary half-peak pattern on the domain $[-A,\Lambda/2]$.
The corresponding changes in the domain length of the elementary patterns do not affect the peak profile within the sharp-peak approximation.
However, the length changes do affect the plateau profile.
Generalizing the plateau profile, Eq.~\eqref{eq:coal-tail-approx}, for a shifted peak positioned at a distance ${L_- = \Lambda/2+A}$ from the boundary at ${x = \Lambda/2}$, one obtains
\begin{equation}
    \cStat^\mathrm{plateau}(x; L_-) 
    = c_- + 2 a_- \exp\left[-\frac{L_-}{\ell_-}\right]
    \cosh  \left[\frac{{\Lambda}/{2} - x}{\ell_-}\right].
\end{equation}
The coalescence mode in the right plateau $\delta c_\mathrm{plateau}$ can hence be approximated by the difference between the stationary plateau profiles of the shifted and unshifted peaks
\begin{align}
    \delta c_\mathrm{plateau}(x) &\approx 
    \cStat^\mathrm{plateau}(x; \Lambda/2+A)-\cStat^\mathrm{plateau}(x; \Lambda/2)\nonumber\\
    &\approx 
    A \, \partial_{L_-}\cStat^\mathrm{plateau}(x; L_-){\big|}_{L_-=\Lambda/2}\nonumber\\
    &= -\frac{\delta c_-}{\ell_-} \cosh \bigg[\frac{\Lambda/2-x}{\ell_-} \bigg]
    \, ,
\label{eq:delta-c-plateau-approx}
\end{align}
where the approximation in the second line is the lowest-order of the Taylor expansion for small shifts $A$.

Using Eq.~\eqref{eq:delta-c-plateau-approx}, we approximate the change in $\check c$ during coalescence, $\delta c|_{\Lambda/2}$, by the change of the stationary pattern profile $\delta c_\mathrm{stat}$ under a change in the plateau length as
\begin{equation} 
\label{eq:delta-c}
    \delta c |_{\frac{\Lambda}{2}} 
    \approx  
    \delta c_\mathrm{plateau}(\Lambda/2) 
    = 
    - \frac{A}{\ell_-} \, \delta c_-\, .
\end{equation}
Finally, the tail approximation, Eq.~\eqref{eq:coal-tail-approx}, also implies
\begin{equation}\label{eq:tail-curvature-coalescence}
    \partial_x^2\cStat|_{\Lambda/2} 
    \approx 
    \frac{1}{\ell_-^2} \, \delta c_-
    \, .
\end{equation}
Then, inserting Eqs.~\eqref{eq:delta-c},~\eqref{eq:tail-curvature-coalescence} into Eq.~\eqref{eq:coales-sigma-eq-1}, one obtains a closed expression for the growth rate in terms of the stationary pattern profile.
Comparing the terms of this closed-form expression with Eq.~\eqref{eq:peak-eta-m} for the change of the stationary mass-redistribution potential with the domain length, one can write this rate expression more compactly as
\begin{widetext}
\begin{equation} 
\frac{\sigma}{T}\left[\int_0^{\Lambda / 2}\mathrm{d}x\, |\partial_x \cStat|  (\partial_\eta \tilde{f}) \int_0^x\mathrm{d}y \frac{\rhoStat(y)-\rho_-}{\chi_\rho(\rhoStat(y))\rhoStat(y)}  + T \int_0^{\Lambda / 2}\mathrm{d} x\, (\partial_x \cStat)^2  \right]
= \left(\partial_{\Lambda/2}^{} \eta_\mathrm{stat}\right) \Tilde{F}_\eta\, .
\end{equation}
As defined above in Eq.~\eqref{eq:Ftilde}, one has ${\Tilde{F}_\eta = \int_{c_-}^{\hat{c}(M)}\mathrm{d}c\, \partial_\eta \tilde{f}(c,\etaStat^\infty(M))}$.
Solving for $\sigma$, multiplying by ${\chi_\rho^0 \equiv \chi_\rho(\rhoStat(0))}$, 
and using the definition of $\partial_M^-\etaStat$ given in Eq.~\eqref{eq:peak-eta-m} yields
\begin{equation} \label{eq:coales-sigma-eq-2}
\sigma = 
-\frac{2 T \chi_\rho^0 (\hat{\rho}-\check{\rho}) \partial_M^- \eta_\mathrm{stat}}{\frac{1}{\Tilde{F}_\eta} \int_0^{\Lambda / 2}\mathrm{d}x\, |\partial_x \cStat| \chi_\rho^0 (\partial_\eta \tilde{f}) \int_0^x\mathrm{d}y \frac{\rhoStat(y)-\rho_-}{\chi_\rho(\rhoStat(y))\rhoStat(y)} 
    + 
    \frac{T}{\Tilde{F}_\eta} \chi_\rho^0 \int_0^{\Lambda / 2}\mathrm{d} x\, (\partial_x \cStat)^2}\, .
\end{equation}
This is an explicit expression for the growth rate $\sigma$ in terms of the stationary pattern profile within singular perturbation theory.
\end{widetext}

\paragraph{Disentangling the rate expression.\;---}
The remainder of this section is dedicated to the physical interpretation of the different terms.
To this end, we interpret the integral expressions as different averages over the interface (half-peak) region.
Moreover, these averages define different measures for the interface or half-peak width.
Using the averages and widths, the growth rate can be expressed in terms of coarse-grained properties of the peak.

Thus, we define a distribution function $P(x)$ which is localized to the interface region by choosing it to be proportional to the gradients $\partial_x \cStat$ of the stationary chemoattractant profile
\begin{equation} 
    P(x) 
    \equiv 
    \frac{|\partial_x \cStat|}{\hat{c}-\check{c}}.
\end{equation}
The denominator is chosen such that the distribution function is (approximately) normalized when integrated over an elementary (half-peak) pattern.
Thus, 
\begin{equation} 
\label{eq:int-avg-translMode}
    \langle \bullet\rangle_\mathrm{int} 
    \equiv 
    \int_0^{\frac{\Lambda}{2}} \mathrm{d}x\, P(x) \, \bullet
\end{equation}
defines a weighted average over one interface of the (symmetric) peak.
As the average over the magnitude of the distribution $\langle P(x)\rangle_\mathrm{int}$ corresponds to the average height of the distribution, and $P(x)$ is normalized, a width $\ell_\mathrm{int}$ of the distribution can be defined by distributing the area under the curve into a rectangle such that
\begin{equation}\label{eq:int-width-1}
    \ell_\mathrm{int}\equiv \frac{1}{\langle P(x)\rangle_\mathrm{int}}.
\end{equation}
This expression is inversely proportional to the second integral in the denominator of the growth rate Eq.~\eqref{eq:coales-sigma-eq-2}, giving an interpretation of this integral.

Similarly, also the distribution function
\begin{equation}
    \tilde P (x) 
    \equiv 
    \frac{1}{\Tilde{F}_\eta} \,  (\partial_\eta \tilde{f}) \, 
    |\partial_x \cStat|
\end{equation}
is localized at the interface.
Again, the normalization is chosen such that the distribution is (approximately) normalized when integrated over an elementary pattern.
The reaction rate $\tilde f_\eta$ reweighs different locations within the interface region based on how strongly changes of $\eta$ change to the chemoattractant reactions.
Using the distribution $\tilde P(x)$, we define the second interface average
\begin{equation} 
\label{eq:int-avg-2}
    \langle \bullet \rangle_{\widetilde {\mathrm{int}}} 
    \equiv 
    \int_0^{\Lambda/2} \mathrm{d}x\, \tilde P(x) \, \bullet
    \, .
\end{equation}
This average gives an interpretation to the first integral in the denominator of the growth rate Eq.~\eqref{eq:coales-sigma-eq-2}.

The quantity averaged in this first integral in the denominator of the coalescence growth rate can be interpreted as follows.
We define
\begin{equation} 
    \tilde{L}(x) 
    \equiv
    \int_0^x\mathrm{d}y \, 
    \chi_\rho^0 \, 
    \frac{\rhoStat - \check\rho}{\chi_\rho \rhoStat}.
\end{equation}
Within the peak region, the integrand fulfills
\begin{equation} 
    \frac{\chi_\rho^0}{\chi_\rho}\left(1-\frac{\check\rho}{\rhoStat}\right) \sim 1\, ,
\end{equation}
because one has ${\check\rho \ll \rhoStat}$ for large peaks in the peak region and close to $x=0$ we expect ${\chi_\rho^0/\chi_\rho = \mathcal{O}(1)}$.
In this sense, the integral $\Tilde{L}(x)$ measures a distance from ${x = 0}$.\footnote{
Considering variations in $\chi_\rho^0/\chi_\rho$, one can see that a decrease in $\chi_\rho$ compared to $\chi_\rho^0$ leads to an increased ``length'' $\Tilde{L}$.
As $\Tilde{L}$ will turn out to measure the length scale of the gradient in the mass-redistribution potential between the two peak interfaces during diffusion-limited peak coalescence, this behavior implies that a reduced mobility $\chi_\rho$ [cf.\ Eq.~\eqref{eq:cont-eq}] away from the peak maximum leads to an increased effective interface width $\Tilde{\ell}_\mathrm{int}$ (defined below) that suppresses the peak-coalescence rate by increasing the effective distance of mass transport.}
Thus, with the second weighted average Eq.~\eqref{eq:int-avg-2} the quantity
\begin{equation}\label{eq:int-width-2}
    \Tilde{\ell}_\mathrm{int} \equiv \langle\Tilde{L}(x)\rangle_{\widetilde{\mathrm{int}}}
\end{equation}
has units of a length and gives a second measure for the interface (half-peak) width.

\medskip

\paragraph{The coalescence growth rate.\;---}
Inserting the definitions of the interface widths Eqs.~\eqref{eq:int-width-1},~\eqref{eq:int-width-2} and the definition of the interface average Eq.~\eqref{eq:int-avg-translMode} into Eq.~\eqref{eq:coales-sigma-eq-2} yields
\begin{equation} 
\label{eq:sigma-coal}
    \sigma^- 
    = 
    -\frac{2 \partial_M^- \eta_\mathrm{stat} }{\frac{\tilde \ell_\mathrm{int}}{ T \chi_\rho^0 \Delta\rho} + \frac{\Delta c}{\Delta\rho \ell_\mathrm{int} \langle \partial_\eta \tilde{f}\rangle_{\mathrm{int}}}}
    \, ,
\end{equation}
with $\Delta\rho \equiv \hat{\rho}-\check{\rho} \approx \hat{\rho}-\rho_-$ and $\Delta c \equiv \hat{c}-\check{c} \approx \hat{c}-c_-$.
Here, we reintroduced the superscript $-$ to denote the coalescence rate and distinguish it from the peak-competition rate calculated in the following section. 
Note that one has ${\langle \partial_\eta \tilde{f}\rangle_{\mathrm{int}} > 0}$, as discussed in Sec.~\ref{sec:stat-patterns}.
Thus, the growth rate $\sigma^-$ is positive because ${\partial_M^-\etaStat < 0}$.
Thus, coalescence occurs because the stationary mass-redistribution potential generically decreases with the plateau lengths.
This follows from the total turnover balance of the chemoattractant in stationary patterns [cf.\ Eq.~\eqref{eq:peak-eta-m}].

As we explain in the following, the two terms in the denominator describe the diffusion- and reaction-limited rates $\sigma^-_\mathrm{D,R}$ given by (cf.\ Ref.~\cite{Weyer.etal2023})
\begin{align}
    \sigma^-_\mathrm{D} 
    &= \frac{2 T \chi_\rho^0 \Delta\rho}{\tilde \ell_\mathrm{int}} \,
    \partial_M^- \eta_\mathrm{stat}
    \, , 
    \\
    \sigma^-_\mathrm{R} 
    &=  \frac{2\Delta\rho \,  \ell_\mathrm{int} \langle \partial_\eta \tilde{f}\rangle_{\mathrm{int}}}{\Delta c} \, 
    \partial_M^- \eta_\mathrm{stat}\, ,
\end{align}
with the full coalescence timescale $1/\sigma^-$ following by adding both timescales:
\begin{equation}\label{eq:coal-rate-timescale-addition}
    \frac{1}{\sigma^-} = \frac{1}{\sigma^-_\mathrm{D}} + \frac{1}{\sigma^-_\mathrm{R}}\, .
\end{equation}

The rate $\sigma^-_\mathrm{D}$ agrees with the QSS result obtained in Ref.~\cite{Weyer.etalinpreparation}.
As discussed therein, this rate describes the redistribution of mass through the peak to enable its movement.
The mathematical derivation performed above provides an explicit definition of the interface or peak (half-)width $\Tilde{\ell}_\mathrm{int}$, Eq.~\eqref{eq:int-width-2}, that determines the effective (half-)length over which mass has to be transported to move the peak and is an undetermined quantity in the QSS derivation of Ref.~\cite{Weyer.etalinpreparation}.

One finds ${\sigma^-\to\sigma^-_\mathrm{D}}$ in the limit of quick local relaxation (fast local reactions ${\langle \partial_\eta \tilde{f}\rangle_{\mathrm{int}}\to\infty}$) or when redistribution is slow (chemotaxis strength ${T\to 0}$).
However, in the full coalescence rate $\sigma^-$, the reaction-limited timescale adds to the timescale of mass redistribution.
Depending on the finite, average reaction rate $\langle \partial_\eta \tilde{f}\rangle_{\mathrm{int}}$, the rate $\sigma^-_\mathrm{R}$ describes the rate of locally (reactively) adapting the chemoattractant density to the peak shift.
This interpretation becomes evident by taking the limit of fast chemotaxis ${T \chi_\rho^0 \Delta\rho/\tilde \ell_\mathrm{int}\to \infty}$ which allows for the instantaneous redistribution of cells (mass) through the peak (${\sigma^-_\mathrm{D}\to\infty}$).
In this limit of instantaneous redistribution of the cell density $\rho$, the chemoattractant dynamics must be rate-limiting.
In this regime, one finds ${\sigma^-\to\sigma^-_\mathrm{R}}$.
The full coalescence process contains both processes:
The chemoattractant density (mass-redistribution potential) must adapt on both sides of the peak to the changed plateau lengths, which then creates the gradient that redistributes cells from one side of the peak to the other.
Thus, one expects the full timescale of coalescence $1/\sigma^-$ to be the sum of the reactive timescale of the regional relaxation $1/\sigma^-_\mathrm{R}$ plus the timescale of the subsequent redistribution of the cells $1/\sigma^-_\mathrm{D}$.
Our singular-perturbation result for the overall cealescence rate Eq.~\eqref{eq:coal-rate-timescale-addition} verifies this reasoning.
A detailed discussion of the interplay of the redistribution and reaction processes is given in Ref.~\cite{Weyer.etal2023} in the context of mass-conserving reaction--diffusion systems.

Taken together, the above singular perturbation calculation gives the rate of the coalescence mode of the mass-competition instability.
This rate depends on the changes $\partial_M^-\etaStat$ of the stationary mass-redistribution potential in the pattern plateaus left and right of the peak that are induced by the peak shift.
In the diffusion-limited regime, the rate follows from calculating the rate of mass transport from the resulting gradient in $\eta$.
In the reaction-limited regime, the rate is set by how strongly the shift $\partial_M^-\etaStat$ changes chemoattractant production and degradation in the interface regions.
The overall timescale of mass competition follows from adding the timescales of both subprocesses.
We have expressed the rate using averages over the stationary peak profile, which can be interpreted as the coarse-grained properties of the aggregates relevant to their dynamics.
These properties can be calculated for specific mathematical models by determining the specific stationary peak profiles.
As an example, we calculate the coalescence rate explicitly for the mKS model in Sec.~\ref{sec:mKS}.

\subsubsection{Peak competition}
\label{sec:peak-comp}

In this section, we perform the singular perturbation analysis of the peak-competition scenario of the mass-competition instability.
The calculation proceeds analogously as for the coalescence mode but the competition mode must be approximated differently from the coalescence mode.

\medskip

\paragraph{Approximation of the competition mode.\;---}
Peak competition is driven by a transfer of mass between two peaks, and we study the neighbor-neighbor interaction by the competition of two half-peaks at the reflective boundaries of the domain ${I = [-\Lambda/2,\Lambda/2]}$ [see Fig.~\ref{fig:massComp-scenarios}(b)].
Again, we assume that the peaks are narrow relative to the domain length $\Lambda$ and that the relaxation of individual peaks to the stationary profile is fast compared to the timescale of mass competition (cf.\ Sec.~\ref{sec:assumptions}).
Thus, we again approximate the peak profiles during mass competition by stationary peak profiles.
Because mass is transferred between the peaks, the left and right peaks have to be approximated by stationary profiles $\rho\sim\rhoStat(x;\etaStat(M\mp A))$ for peak masses $M\mp A$ that differ by a small amount $2 A$.
Linearizing the profile deviation $\rho-\rhoStat(x;\etaStat(M))$ in $A$, we approximate the antisymmetric competition mode by
\begin{equation}
    \delta\rho 
    \approx 
    \pm A \, \frac{\partial_{\etaStat} \rhoStat}{\partial_{\etaStat} M}
\end{equation}
in the right and left peak regions close to the domain boundaries, respectively.
Analogously we approximate ${\delta c \approx \pm A \, \partial_{\etaStat} \cStat/\partial_{\etaStat} M}$.
As a result, we approximate the competition mode in the peak regions by the mass mode $\partial_{\etaStat} \rhoStat(x)$, which was discussed in Sec.~\ref{sec:stat-patterns} to describes the change of the elementary pattern profiles if the mass $M$ of the peak is changed.

As in the case of peak coalescence, the fast relaxation of the plateau regions implies that the mass-redistribution potential fulfills ${\partial_x^2\delta\eta \approx 0}$ between the peaks.
Thus, to fulfill the antisymmetry we approximate $\delta\eta$ between the peaks and within the sharp-peak approximation, i.e., neglecting the width of the peaks positioned at $\pm \Lambda/2$ by [cf.\ Fig.~\ref{fig:massComp-scenarios}(b)]
\begin{equation}\label{eq:comp-eta-grad}
    \delta\eta \approx 2 \delta\eta|_{\Lambda/2} \frac{x}{\Lambda}\, .
\end{equation}
As $\eta$ drives mass redistribution, the amplitude of the gradient $\delta\eta$ must match the rate of mass change in the peaks.
Therefore, the amplitude of $\delta \eta$ can be related to $A$ by integrating the linearized continuity equation Eq.~\eqref{eq:massComp-lin-cont-eq} over one domain half, which yields
\begin{align}
    \sigma A \frac{\partial_{\etaStat} M + \Lambda \partial_{\etaStat} \rho_-}{2 \partial_{\etaStat} M} &\approx \sigma \frac{A}{2} \\
    &\approx -T \chi_\rho^- \rho_- \frac{2\delta\eta|_{\Lambda/2}}{\Lambda}\, .
\end{align}
Here, the growth rate $\sigma$ is the growth rate $\sigma^+$ of peak competition.
As in the previous section, we drop the superscript $``+''$ during the following derivation.
In the first line, we neglect the change in the low-density plateau due to the assumption on the mass-change of the plateau, Eq.~\eqref{eq:plateau-negligible}.
In the second approximation, we used the no-flux boundary condition and neglected corrections due to the exponential pattern tails in the plateau region.

\paragraph{Determining the competition growth rate.\;---} Having derived a relation of the two amplitudes via the linearized continuity equation, we again use the linearized dynamics Eq.~\eqref{eq:massComp-lin-c-eq} of the chemoattractant to determine the growth rate.
Projecting the chemoattractant dynamics onto the mass mode, that is, peak growth and shrinking, one obtains from integrating over the right domain half
\begin{align} 
    \int_0^{\Lambda/2}\mathrm{d}x\, 
    &(\partial_{\etaStat} \cStat) (\partial_\eta \tilde{f}) \delta \eta  - \sigma \int_0^{\Lambda/2}\mathrm{d}x \,  (\partial_{\etaStat} \cStat) \delta c
    \nonumber\\ 
    &= \sigma \int_0^{\Lambda/2}\mathrm{d}x \,  (\partial_{\etaStat} \cStat) \, \mathcal{L} \, \delta c 
    \, .
\end{align}
Applying the sharp-peak approximation and using that $\partial_{\etaStat}\cStat$ is localized to the interface region up to the exponential pattern tails [due to condition Eq.~\eqref{eq:plateau-negligible-c}], one can approximate $\delta\eta$ as constant in the integral localized to the narrow peak region.
Moreover, in the second integral on the left-hand side we can use the approximation of $\delta c \sim \partial_{\etaStat}\cStat$, because the integral is localized to the peak region.
To determine the right-hand side, we note that the stationary profile equation Eq.~\eqref{eq:profile-eq-c} implies
\begin{equation}\label{eq:mass-mode-eq}
    \partial_\eta \tilde{f} = \mathcal{L}\, \partial_{\etaStat}\cStat\, .
\end{equation}
Thus,
integrating by parts twice on the right-hand side, and using Eq.~\eqref{eq:mass-mode-eq}, one obtains, with all but one of the boundary term vanishing exactly due to the no-flux boundary conditions,
\begin{widetext}
\begin{align} \label{eq:comp-c-eq-1}
-\sigma \frac{A \Lambda}{4 T \chi_\rho^- \rho_-} \int_0^{\Lambda/2}\mathrm{d}x\, (\partial_{\etaStat} \cStat) \partial_\eta \tilde{f}  &- \sigma \frac{A}{\partial_{\etaStat} M} \int_0^{\Lambda/2}\mathrm{d}x\, (\partial_{\etaStat} \cStat)^2 \nonumber\\
&\approx
-D_c  (\partial_{\etaStat} \cStat) \partial_x \delta c|_0^{} + \frac{A}{\partial_{\etaStat} M}\int_0^{\Lambda/2}\mathrm{d}x\, (\partial_{\etaStat} \cStat) \partial_\eta \tilde{f}\, .
\end{align}
\end{widetext}
Analogously to the previous section, we define the interface weight
\begin{equation} 
P(x) \equiv \frac{\partial_\eta c}{\int_0^{\Lambda/2}\mathrm{d}x\, \partial_\eta c} = \frac{2 \partial_\eta c}{\partial_\eta M_c} \, ,
\end{equation}
with the chemoattractant mass $M_c$ of the peak [cf.\ Eq.~\eqref{eq:chemoattractant-mass}].
This weight is localized to the peak region due to the assumption Eq.~\eqref{eq:plateau-negligible-c} that changes in the plateau mass are negligible.
With this weight, we define the interface average
\begin{equation} 
    \langle \bullet\rangle_\mathrm{int} 
    \equiv 
    \int_0^{\Lambda/2}\mathrm{d}x\, P(x) \, \bullet 
    \, ,
\end{equation}
and the interface width $\ell_\mathrm{int} \equiv (\langle P(x)\rangle_\mathrm{int})^{-1}$.
With these definitions, we can rewrite the integrals in Eq.~\eqref{eq:comp-c-eq-1}, which yields
\begin{align} \label{eq:comp-c-eq-2}
    &-\sigma \frac{A \Lambda}{8 T \chi_\rho^- \rho_-} \langle\partial_\eta \tilde{f}\rangle_\mathrm{int} \partial_{\etaStat} M_c   
    - \sigma \frac{A}{\partial_{\etaStat} M} \frac{(\partial_{\etaStat} M_c)^2}{4 \ell_\mathrm{int}} 
    \nonumber\\
    &\qquad\approx
    -D_c  \, (\partial_{\etaStat} \cStat) \, \partial_x \delta c|_0^{} 
    \nonumber\\
    &\quad \qquad \, + 
    \frac{A}{\partial_{\etaStat} M} \, \langle\partial_\eta \tilde{f}\rangle_\mathrm{int} \,  \frac{\partial_{\etaStat} M_c}{2}
    \, .
\end{align}
Finally, we have to estimate the boundary term on the right-hand side because it is not yet formulated in terms of the stationary pattern profile.
To this end, we estimate profile $\delta c$ in the plateaus in terms of the gradient $\delta\eta$.
Because the plateau densities relax quickly, we approximate the gradient in $\delta\eta$ as linear in the plateau [cf.\ Eq.~\eqref{eq:comp-eta-grad}].
In particular, mass competition is slow compared to the local reactive relaxation of the plateau densities, that is, we have $\tilde{f}^-_{c,\eta}\gg\sigma$.
Thus, up to corrections of the size $\sigma/\tilde{f}^-_c\ll 1$ and the exponential pattern tails, the linearized dynamics for the chemoattractant Eq.~\eqref{eq:massComp-lin-c-eq-1} is solved in the plateaus by $\delta c \approx \delta\eta/\partial_c\eta^*(c_-)$ [cf.\ Eq.~\eqref{eq:nullcline-restriction-plateaus}].
As a result, in the plateaus, the densities are restricted to the nullcline, that is, one has $\eta\approx\eta^*(c)$.
Thus, neglecting the contributions from the pattern tails exponentially small in the domain length $\Lambda$, one has
\begin{align}
    -D_c  (\partial_{\etaStat} \cStat) \partial_x \delta c|_0^{}&\approx -D_c (\partial_{\etaStat}c_-)^2 \partial_x\delta\eta|_0 \nonumber\\
    &\approx \sigma \frac{D_c A}{2 T \chi_\rho^- \rho_-} (\partial_{\etaStat}c_-)^2\, .
\end{align}
As we assume changes in the low-density plateau density $c_-$ to be small [${\Lambda \partial_{M_c}c_- \ll 1}$, cf.\ Eq.~\eqref{eq:plateau-negligible-c}], this shows that the boundary term can be neglected compared to the other terms scaling $\sim \partial_{\etaStat}M_c$.

\paragraph{The competition growth rate.\;---}
With these approximations, Eq.~\eqref{eq:comp-c-eq-2} yields the growth rate of the peak-competition mode
\begin{equation}\label{eq:sigma-comp}
    \sigma^+ = - \frac{\partial_M^{} \etaStat}{\frac{\Lambda}{4 T \chi_\rho^- \rho_-} + \frac{\partial_M^{} M_c}{2 \ell_\mathrm{int}\langle\partial_\eta \tilde{f}\rangle_\mathrm{int}}}\, .
\end{equation}
Note that we assume ${\partial_M^{}M_c>0}$ [cf.\ Eq.~\eqref{eq:assumption-chemoattrMass-increase}] such that the negative sign of $\partial_M^{}\etaStat$ results in a positive growth rate.
As for coalescence, the rate for peak competition can be decomposed into a diffusion-limited rate $\sigma^+_\mathrm{D}$ that agrees with the QSS rate derived in Ref.~\cite{Weyer.etalinpreparation} and a reaction-limited rate $\sigma^+_\mathrm{R}$ as $(\sigma^+)^{-1} = (\sigma^+_\mathrm{D})^{-1} + (\sigma^+_\mathrm{R})^{-1}$.
The full rate $\sigma^+$ is positive because ${\partial_M^{}\etaStat < 0}$ (see Sec.~\ref{sec:stat-patterns}).

Taken together, peak competition is driven by the same processes of mass redistribution between and reactive relaxation at the peaks as peak coalescence.
However, while peak coalescence is due to mass competition by self-amplifying gradients between the shortening and elongating plateaus, peak competition is driven by mass competition between a growing and a shrinking peak.

\subsubsection{Mass-competition rates for mesa patterns}

For mesa patterns, the same derivation can be employed (cf.\ Ref.~\cite{Weyer.etal2023} for the analysis of mesa patterns in two-component mass-conserving reaction--diffusion systems).
Again, the profile $\delta\eta$ can be approximated as straight (${\partial_x^2\eta \approx 0}$) in the plateaus.
Mass is redistributed from one interface to the other and the amplitude of the induced $\eta$ gradient between the interfaces depends on the width of the internal plateau length separating the two competing interfaces.
Within the sharp-interface approximation, the half-lengths of the low- and high-density plateaus are [see Eq.~\eqref{eq:plateau-lengths}]
\begin{equation}
    L_\pm 
    \equiv 
    \xi_\pm \, \frac{\Lambda}{2}.
\end{equation}
With this, one obtains the competition rate for two high-density (half) mesas [$\rhoStat(\pm\Lambda/2)\approx\rho_+$]
\begin{equation} 
\label{eq:sigma-comp-mesa}
    \sigma^+ 
    = -\frac{2 \, \partial_M^+ \eta_\mathrm{stat} }{\frac{\xi_- \Lambda}{ 2 T \chi_\rho^- \rho_-} + \frac{\Delta c}{\Delta\rho \ell_\mathrm{int} \langle \partial_\eta \tilde{f}\rangle_{\mathrm{int}}}}\, .
\end{equation}
The interface average $\langle\bullet\rangle_\mathrm{int}$ and the interface width $\ell_\mathrm{int}$ are defined with respect to the distribution ${P(x) = \partial_x\rhoStat/\Delta\rho}$ [cf.\ Eqs.~\eqref{eq:int-avg-translMode},~\eqref{eq:int-width-1}].

The mass-competition rate for mesa coalescence analogously reads
\begin{equation} 
\label{eq:sigma-coal-mesa}
\sigma^- = -\frac{2 \partial_M^- \eta_\mathrm{stat} }{\frac{\xi_+ \Lambda}{ 2 T \chi_\rho^+ \rho_+} + \frac{\Delta c}{\Delta\rho \ell_\mathrm{int} \langle \partial_\eta \tilde{f}\rangle_{\mathrm{int}}}}\, .
\end{equation}

Similar as for peak patterns, in mesa patterns, mass competition between interfaces drives both the competition high-density mesas and their coalescence.
While the competition process is driven by length changes of the high-density plateaus, that is, $\partial_M^+\etaStat$, coalescence depends on the length changes of the low-density plateaus, that is, $\partial_M^-\etaStat$.
Again, the growth rates of the mass-competition instability are given in terms of collective properties of the pattern interfaces which are determined as averages over the interface and their positions.

\section{Mass competition and coarsening in the minimal Keller-Segel Model}
\label{sec:mKS}
As an example, we calculate the growth rate for neighbor-neighbor peak coalescence for the minimal Keller--Segel model using the stationary peak profiles Eq.~\eqref{eq:mKS-stat-rho}--\eqref{eq:mKS-stat-c-outer} derived in Ref.~\cite{Kang.etal2007}.
In particular, we show that the resulting growth rate for peak coalescence agrees with their result, Eq.~2.20, for $\Lambda\gg 1$.

In the minimal Keller-Segel model, ${\chi_c = \chi_\rho = 1}$, and ${f = \rho -c}$; see Eqs.~\eqref{eq:mKS}.
To calculate the growth rate of peak coalescence Eq.~\eqref{eq:sigma-coal}, we first determine the individual quantities necessary.
We assume a large domain ${\Lambda \gg \ell_\mathrm{int}}$ such that integrals over peak-localized quantities on the interval ${I = [0,\Lambda/2]}$ can be replaced by integrals over the stationary pattern on the half-infinite line $[0, \infty)$.
The cell-density and chemoattractant profiles Eq.~\eqref{eq:mKS-stat-rho},~\eqref{eq:mKS-stat-c-inner},~\eqref{eq:mKS-stat-c-outer} give
\begin{subequations}
\begin{align}
    \Delta \rho 
    &\approx \hat{\rho} = \frac{M^2 T}{8 D_\rho D_c}
    \, ,\\
    \Delta c 
    &\approx \hat{c} = \frac{M}{2 \sqrt{D_c}}
    \, .
\end{align}
\end{subequations}
Moreover, the chemoattractant profile, Eq.~\eqref{eq:mKS-stat-c-outer}, determines the first interface width via its definition, Eq.~\eqref{eq:int-width-1}, as
\begin{equation}
    \ell_\mathrm{int} 
    \approx 
    \frac{\Delta c^2}{\int_0^\infty\mathrm{d}x \, (\partial_x\cStat^\infty(x))^2}  
    = 
    2 \sqrt{D_c}\, .
\end{equation}
The contribution from the inner peak region ${x \sim 1/M}$, in which the chemoattractant profile must be approximated by Eq.~\eqref{eq:mKS-stat-c-inner}, not Eq.~\eqref{eq:mKS-stat-c-outer}, can be neglected in the limit of large peak masses because it only accounts for the peak within a region of width $\sim 1/M$.
Thus, it contributes only a correction $\sim 1/M$ to the interface (half-peak) width $\ell_\mathrm{int}$.
Next, the average reaction rate becomes by using the relations Eqs.~\eqref{eq:density-expressions} between $\rho$, $c$, and $\eta$ and the definition Eq.~\eqref{eq:int-avg-translMode} of the interface average
\begin{align}
    \langle\partial_\eta \tilde{f}\rangle_\mathrm{int} &= -\frac{1}{\Delta c} \int_0^\infty\mathrm{d}x\, (\partial_x\cStat) \partial_\eta \tilde{f} \nonumber\\
    &= -\frac{1}{\Delta c} \int_0^\infty\mathrm{d}x\, (\partial_x\rhoStat) \frac{\partial_\rho f}{\chi_c} \nonumber\\
    &\approx \frac{\hat{\rho}}{\hat{c}} = \frac{M T}{4 D_\rho \sqrt{D_c}}\, .
\end{align}
Furthermore, the second interface width Eq.~\eqref{eq:int-width-2} becomes
\begin{align}
    \Tilde{\ell}_\mathrm{int} &\approx -\frac{1}{\hat\rho} \int_0^\infty\mathrm{d}x\, (\partial_x \rhoStat^\infty) \int_0^x\mathrm{d}y \left(1-\frac{\rho_-}{\rhoStat^\infty}\right) \nonumber\\
    &= \int_0^\infty\mathrm{d}x\frac{(\rhoStat^\infty-\rho_-)^2}{\hat{\rho} \rhoStat^\infty} \nonumber\\
    &\approx \frac{M}{2\hat{\rho}} = \frac{4 D_\rho D_c}{M T}\, .
\end{align}
In order to obtain $\partial_M^- \eta_\mathrm{stat} \propto \partial_{\Lambda/2} \eta_\mathrm{stat}$, we determine $\delta c_\mathrm{stat}$ using the finite-domain approximation given in Ref.~\cite{Kang.etal2007} at the boundary.
This reads
\begin{align} 
\delta c_\mathrm{stat}|_\frac{\Lambda}{2} &\approx \left.\frac{M}{\sqrt{D_c}}\frac{\cosh\left(\frac{\Lambda}{2\sqrt{D_c}}\right)}{\sinh\left(\frac{\Lambda}{\sqrt{D_c}}\right)} \cosh\left(\frac{x - \frac{\Lambda}{2}}{\sqrt{D_c}}\right)\right|_\frac{\Lambda}{2} \\
&\approx \frac{M}{2\sqrt{D_c}} e^{\frac{-\Lambda}{2\sqrt{D_c}}}  \overset{!}{=} 2 a_-^c e^{-\frac{\Lambda}{2\ell_-}}\, ,
\end{align}
where we compare the expression to the general expression of the pattern tail, Eq.~\eqref{eq:coal-tail-approx}, in the last line.
From this, we find $a_-^c = M/(4\sqrt{D_c})$ and $\ell_- = \sqrt{D_c}$ and obtain from Eq.~\eqref{eq:peak-eta-m}
\begin{equation} 
\partial_M^- \eta_\mathrm{stat} \approx \frac{32 D_\rho^2 \sqrt{D_c}}{M^2 T^2} \exp\left(-\frac{\Lambda}{\sqrt{D_c}}\right) ,
\end{equation}
such that one finds
\begin{equation} \label{eq:mKS-sigma-coal}
\sigma^- \approx \frac{2 \frac{T M }{D_c^{3/2}}}{1 + \frac{T M}{4 D_c^{3/2}}} \exp\left(-\frac{\Lambda}{\sqrt{D_c}}\right)\, ,
\end{equation}
in accordance with Kang et al.\ \cite{Kang.etal2007}.

The rate is exponentially small in the wavelength $\Lambda$ because a shift of the peak only changes the widths of the left and right plateau, which has an effect exponentially small in the plateau width on the total turnover balance Eq.~\eqref{eq:ttb} and thus $\etaStat$.
Interestingly, the prefactor shows that with increasing the peak mass $M$ the rate approaches the reaction-limited regime.
In contrast, in peak competition, one expects that the growth rate of the mass-competition instability typically approaches the diffusion-limited regime as the peak mass $M$ increases. 
The reason is that, as $M$ grows, mass must be redistributed between the peaks over increasingly long distances ${\Lambda \approx M/\bar{\rho}}$.
The last approximation holds because ${\rho_- \approx 0}$ for the mKS model [cf.\ Eq.~\eqref{eq:peak-mass-rhobar}].
Thus, mass transport between peaks becomes slow.\footnote{
If the reaction-limited growth rate decreases more strongly with the peak mass $M$, that is, if one has ${\partial_M M_c \chi_\rho^-\rho_-/(\ell_\mathrm{int}\langle \partial_\eta \tilde{f}\rangle_\mathrm{int})\gg M}$, also peak competition approaches the reaction-limited regime for large peak masses.
However, this condition is more stringent than for peak coalescence, for which the reaction-limited regime is approached for large peak masses if ${\Delta c \chi_\rho^+/(\ell_\mathrm{int}\tilde{\ell}_\mathrm{int}\langle \partial_\eta \tilde{f}\rangle_\mathrm{int})}$ increases with the peak mass.}
For peak coalescence, however, mass is transported only through the peak itself, whose width does not grow strongly (here, it even decreases) with the peak mass.
As a result, in the mKS model, the rate of coarsening in the long-time limit is determined by the reaction-limited rate if coalescence is faster than competition (see Ref.~\cite{Weyer.etalinpreparation}).

\section{Conclusions}
\label{sec:conclusions}

Keller--Segel models describe the formation of chemotactic aggregates.
Here, we have analyzed the generic dynamics of these chemotactic aggregates.
In Ref.~\cite{Weyer.etalinpreparation}, we argued within a quasi-steady-state (QSS) approximation that the formation of aggregates from perturbations around the homogeneous steady state is described by a nullcline-slope criterion.
Moreover, within this approximation, we have shown that chemotactic aggregates generically undergo coarsening due to the coalescence of aggregates and their competition for cells~\cite{Weyer.etalinpreparation}.
Here, we have complemented the QSS analysis with the complete linear stability analysis of the homogeneous steady state and a singular perturbation analysis of the coarsening process.

The linear stability analysis, as presented in Sec.~\ref{sec:lsa}, shows that the lateral instability of the homogeneous steady state in Keller--Segel models is always a long-wavelength instability and is exactly predicted by the nullcline-slope criterion,  Eq.~\eqref{eq:nullcline-slope-crit}. This criterion is analogous to the curvature criterion for spinodal decomposition in equilibrium phase-separating systems (comparing the mass-redistribution potential with the chemical potential ${\mu = \delta \mathcal{F}/\delta\phi}$) \cite{Doi2013} and the mass-redistribution instability in two-component mass-conserving reaction--diffusion systems \cite{Brauns.etal2020}.

Subsequently in Sec.~\ref{sec:stat-patterns}, we constructed the fully nonlinear stationary patterns, which can take the form of either peak- or mesa-shaped profiles.
We then investigated the stability of these periodic stationary patterns under the condition that the peaks and mesas are narrow compared to their separation (sharp-interface/sharp-peak approximation).
Our analysis revealed that small mass imbalances between neighboring peaks lead to self-amplifying mass transport from the smaller toward the larger peak, ultimately causing the collapse of the smaller peak.
Analogously, a peak positioned, for instance, closer to its right neighbor than to its left will continue to shift toward its closer right neighbor until coalescence occurs.
This process is driven by self-amplifying mass transport from one peak interface to the other.
Both processes induce the uninterrupted coarsening of patterns with several peaks (and analogously mesas) by the successive collapse and merging of peaks (mesas).
We call these two processes the peak-competition and peak-coalescence scenarios of the mass-competition instability.
The singular perturbation analysis provides the growth rates of these mass-competition instabilities.

Using a QSS analysis, we also identified a mass-competition instability in Ref.~\cite{Weyer.etalinpreparation}, driven by self-amplifying mass transport.
The singular perturbation analysis additionally shows that the mass-competition process is limited by two timescales.
First, mass must be transported from one peak to the other or throughout the peak.
This process resembles the competition process in Cahn--Hilliard systems \cite{Cahn.Hilliard1958} in which gradients in the chemical potential, instead of the mass-redistribution potential drive mass transport and coarsening (cf.\ Ref.~\cite{Weyer.etal2023}).
Because the chemical potential is a functional of the density profile in Cahn--Hilliard systems, redistribution of the density is the only process determining the rate of mass-competition.
In contrast, in the Keller--Segel models (and two-component mass-conserving reaction--diffusion systems as well \cite{Weyer.etal2023}), the mass-redistribution potential has a dynamic equation itself and an additional reactive timescale arises.
Namely, the chemoattractant profile must adapt to the changing peak mass or shifting peak positions.
This adaptation is limited by the reaction rates of chemoattractant production and degradation.
The resulting competition process is analogous to the competition in conserved Allen--Cahn systems \cite{Rubinstein.Sternberg1992}.
In conserved Allen--Cahn systems, mass transport occurs instantaneously because the total mass is only conserved globally but a timescale is associated with the incorporation of mass at the pattern interfaces (cf.\ Ref.~\cite{Weyer.etal2023}).
Our perturbation analysis shows that, as both the redistribution and reactive-relaxation processes occur subsequently, in the Keller--Segel models the timescale (inverse growth rate) of the mass-competition process is given as the sum of the diffusive and reactive timescales.
The two different limits of diffusion- and reaction-limited dynamics has been discussed in a concrete Keller--Segel model derived from a particle-based chemotaxis model in Ref.~\cite{Meyer.etal2014}.
Using a quasi-stationary chemoattractant field, Cahn--Hilliard-like macroscopic dynamics was derived starting from a particle-based chemotaxis model in Refs.~\cite{OByrne.Tailleur2020,Dinelli.etal2024}.

Taken together, our results show that uninterrupted coarsening is generic for chemotactic aggregates described by Keller--Segel models, uncovering the mechanisms underlying coarsening described before in concrete systems \cite{Dolak.Schmeiser2005,Potapov.Hillen2005,Kang.etal2007,Cotton.etal2022,Meyer.etal2014,OByrne.Tailleur2020,Dinelli.etal2024}.
Importantly, using the growth rates, the coarsening laws for patterns in large systems with many peaks and mesas can be predicted \cite{Langer1971,Glasner.Witelski2003,Brauns.etal2021,Weyer.etal2023,Weyer.etalinpreparation}.
It will be interesting to analyze coarsening and its scaling laws in chemotactic colloidal systems \cite{Stark2018,Liebchen.Lowen2018} or colonies of engineered bacteria \cite{Curatolo.etal2020}.

Interestingly, these results generalize motility-induced phase separation (MIPS) \cite{Cates.Tailleur2015}.
MIPS can be derived for quorum-sensing particles, which gives rise to chemotaxis-like terms, if the signaling is instantaneous \cite{Tailleur.Cates2008,Fu.etal2012}.
It is found that the coarse-grained system follows Cahn--Hilliard (Model B) dynamics.
Our results suggest that including the dynamics of the signaling molecules, a second dynamical regime occurs dominated by reactive dynamics.
It will be interesting to analyze the noise statistics in this more general setting.

In Ref.~\cite{Weyer.etalinpreparation}, we have shown that the interplay of chemotactic aggregation with cell growth and death induces intricate patterning by interrupting and reversing coarsening.
Together, both processes can induce sustained spatiotemporal dynamics.
Because the arrest of coarsening is due to a modification of the mass-competition process, the mass-competition rates derived here can be extended to understand these more complicated pattern dynamics.
In particular, an extended singular perturbation analysis can be performed including weak cell growth and death as has been done for two-component reaction--diffusion systems \cite{Weyer.etal2023,Kolokolnikov.etal2014}.
Our results on the generic behavior of Keller--Segel models give a basis to analyze biological chemotaxis systems with more complex signaling dynamics \cite{Ziepke.etal2022} and several cell and chemoattractant species \cite{Wolansky2002,Liu.etal2019a,Muramatsu.etalsubmitted}, relevant for instance in microbial systems and the immune system.

The dynamics found here are remarkably similar to the findings in two-component mass-conserving reaction--diffusion systems \cite{Brauns.etal2021,Weyer.etal2023}.
Intriguingly, in these systems, the sum of the two density fields is conserved by restricting the reaction term to only convert particles between the two states.
In contrast, here, the cell density itself is conserved.
This conserved density is then coupled to the non-conserved chemoattractant field.
It is an interesting question under which conditions more broadly two-component systems with one conservation law generically undergo uninterrupted coarsening with both diffusion- and reaction-limited regimes.

\smallskip

\begin{acknowledgments}
We thank Fridtjof Brauns, Natan Dominko Kobilica, and Florian Raßhofer for inspiring discussions.
This work was funded by the Deutsche Forschungsgemeinschaft (DFG, German Research Foundation) through the Excellence Cluster ORIGINS under Germany’s Excellence Strategy – EXC-2094 – 390783311, the European Union (ERC, CellGeom, project number 101097810), and the Chan-Zuckerberg Initiative (CZI).
\end{acknowledgments}

\appendix

\section{Stability of elementary peak patterns}
\label{sec:elementary-stab}
We follow the discussion in Ref.~\cite{Weyer.etal2023} and calculate the relaxation rates for the redistribution of mass between the peak and the plateau of a single elementary pattern.
As these calculations are analogous to the derivation of the rates of the mass-competition instability, we explain the derivation in detail there (Secs.~\ref{sec:peak-coal},~\ref{sec:peak-comp}) and only state the necessary steps here.

To determine the relaxation rate, we consider the linearized dynamics, Eq.~\eqref{eq:massComp-eigenvalue-problem}, for a half-peak positioned on the left boundary ${x = 0}$ with no-flux boundary conditions that extends to ${x = \Lambda/2}$ (linear dynamics around a single elementary stationary pattern).
We denote the relaxation rate by $\sigma_\mathrm{relax}$ and assume it is negative, i.e., the elementary pattern is stable.
We will determine afterward under which condition this assumption holds.

In the limit of a large plateau length, the redistribution of mass between the plateau and the peak region is slow compared to the local relaxation of the densities in the plateau onto the reactive equilibria ${f \approx 0}$.
This follows from Eq.~\eqref{eq:massComp-lin-c-eq-1} by neglecting the term proportional to the relaxation rate ${\sim\sigma_\mathrm{relax}}$ and assuming that gradients are weak, as will be verified below.
With these approximations, Eq.~\eqref{eq:massComp-lin-c-eq-1} reduces to
\begin{equation}\label{eq:nullcline-restriction-plateaus}
    \delta \eta \approx - \frac{\partial_c \tilde{f}|_{c_-}}{\partial_\eta \tilde{f}|_{c_-}}\delta c = \delta c\, \partial_c \eta^*(c)|_{c_-}\, ,
\end{equation}
where we neglected the exponential pattern tails in the plateau region.
Thus, the plateau densities follow the nullcline $\eta^*(c)$ [or equivalently $\eta^*(\rho)$].
Inserting the relation $\delta\eta = [\partial_\rho\eta^*(\rho_-)]\delta\rho$ into the linearized continuity equation Eq.~\eqref{eq:massComp-lin-cont-eq}, it turns into the diffusion equation
\begin{equation}
    \sigma_\mathrm{relax} \delta\rho = T\chi_\rho^-\rho_- [\partial_\rho\eta^*(\rho_-)]\partial_x^2\delta\rho\, .
\end{equation}
If the mode is stable, i.e., $\sigma_\mathrm{relax}<0$, the mode profile in the plateau is given by
\begin{align}\label{eq:relaxMode}
    \delta\eta_\mathrm{plateau}&\propto\delta\rho_\mathrm{plateau}\nonumber\\
    &\propto \cos\left[\sqrt{\frac{-\sigma_\mathrm{relax}}{T\chi_\rho^-\rho_- \partial_\rho\eta^*(\rho_-)}} \left(\frac{\Lambda}{2}-x\right)\right].
\end{align}
Using that also $\delta c$ shows this profile [cf.\ Eq.~\eqref{eq:massComp-delta-c}] and inserting this result in Eq.~\eqref{eq:massComp-lin-c-eq-1}, it becomes clear that the gradient term is small, ${\partial_x^2\delta c\sim \sigma_\mathrm{relax}}$, and can be neglected, as we assumed to arrive at Eq.~\eqref{eq:nullcline-restriction-plateaus}.

From Eq.~\eqref{eq:relaxMode}, one obtains
\begin{align}\label{eq:relax-fraction-outer}
    \frac{\partial_x \delta\eta_\mathrm{plateau}|_{x=0}}{\delta\eta_\mathrm{plateau}|_{x=0}} = \frac{2 \zeta}{\Lambda}\tan(\zeta)\, ,
\end{align}
with $\zeta = \frac{\Lambda}{2}\sqrt{|\sigma_\mathrm{relax}|/[T\chi_\rho^-\rho_- \partial_\rho\eta^*(\rho_-)]}$.
We have neglected that the peak is positioned around ${x = 0}$ for this approximation based on the plateau profiles.

Because we describe the relaxation mode redistributing mass between the peak and the plateau, in the peak region, we approximate the relaxation mode by the mass mode $[\delta \rho \approx \delta M (\partial_{\etaStat}\rhoStat)/\partial_{\etaStat}M, \etaStat \approx\mathrm{const.}]$ that describes the change of the stationary profile under the change of the peak mass.
This ansatz assumes that the relaxation of the peak profile is fast compared to its mass change due to the redistribution of mass between the peak and the plateau.
Integration of the linearized continuity equation Eq.~\eqref{eq:massComp-lin-cont-eq} on the interval $[0,b]$ with $\ell_\mathrm{int}\ll b\ll \Lambda/2$, denoting the interface or half-peak width by $\ell_\mathrm{int}$, yields
\begin{equation}\label{eq:relax-peak-cont-eq}
    \sigma_\mathrm{relax} \frac{\delta M}{2} \approx T\chi_\rho^-\rho_- \partial_x\delta\eta|_{x=b}\, .
\end{equation}
Here, we neglected the exponential pattern tails at $x=b$ and anticipated ${-\partial_{\etaStat}M \gg \Lambda \partial_{\etaStat}\rho_-}$, i.e., that the change in the plateau mass is negligible compared to the change in the peak mass.
The following calculation performed in this section will show that this separation of scales results from the assumption that mass competition is slow compared to the relaxation modes calculated here (cf.\ Sec.~\ref{sec:assumptions}).

Following the same steps as in the calculations below for the peak-competition scenario [Eqs.~\eqref{eq:comp-c-eq-1}], one obtains from integrating Eq.~\eqref{eq:massComp-lin-c-eq}
\begin{widetext}
\begin{equation}
    \int_0^{b}\mathrm{d}x\, (\partial_{\etaStat} \cStat) (\partial_\eta \tilde{f}) \delta \eta  - \sigma \int_0^{b}\mathrm{d}x \,  (\partial_{\etaStat} \cStat) \delta c 
= \int_0^{b}\mathrm{d}x \,  (\partial_{\etaStat} \cStat) \mathcal{L}\delta c\, ,
\end{equation}
which yields
\begin{equation}
\frac{\delta \eta|_{x=0}}{2} \langle \partial_\eta \tilde{f} \rangle_\mathrm{int} (\partial_{\etaStat} M_c) - \sigma \delta M\frac{(\partial_{\etaStat} M_c)^2}{4 (\partial_{\etaStat} M) \ell_\mathrm{int}} = -D_c (\partial_{\etaStat} \cStat) \partial_x\delta c|_{x=b} +\delta M\frac{\partial_{\etaStat} M_c}{2\partial_{\etaStat} M} \langle \partial_\eta \tilde{f} \rangle_\mathrm{int} \, \label{eq:relax-c-eq-1},
\end{equation}
\end{widetext}
where we again neglected the exponential tails at $x=b$ and introduced the definitions (explained in the context of mass competition in Secs.~\ref{sec:peak-coal},~\ref{sec:peak-comp})
\begin{align}
    \langle \partial_\eta \tilde{f} \rangle_\mathrm{int} &\equiv \frac{1}{\partial_{\etaStat} M_c }\int_0^{b}\mathrm{d}x\, (\partial_{\etaStat} \cStat) \partial_\eta \tilde{f}\, ,\\
    M_c &\equiv \int_0^{b}\mathrm{d}x\, (\partial_{\etaStat} \cStat) \, ,\\
    \ell_\mathrm{int} &\equiv \frac{1}{(\partial_{\etaStat} M_c)^2 }\int_0^{b}\mathrm{d}x \,  (\partial_{\etaStat} \cStat)^2.
\end{align}
The boundary term in Eq.~\eqref{eq:relax-c-eq-1} can be neglected because we demand $\Lambda \partial_{M_c}c_-\ll 1$ [see Eq.~\eqref{eq:plateau-negligible-c}].

Combining Eqs.~\eqref{eq:relax-peak-cont-eq},~\eqref{eq:relax-c-eq-1}, one obtains
\begin{equation}\label{eq:relax-fraction-inner}
    \frac{\partial_x\delta\eta|_{x=b}}{\delta\eta|_{x=b}} \approx \frac{\partial_x\delta\eta|_{x=b}}{\delta\eta|_{x=0}} \approx \frac{2 \zeta^2}{\Lambda}\left[ \frac{\Lambda\partial_{\etaStat} \rho_-}{-\partial_{\etaStat}M} + \frac{\sigma_\mathrm{D}^+}{\sigma_\mathrm{R}^+} \zeta^2\right]^{-1}.
\end{equation}
The first approximation follows from $\delta\eta$ being approximately constant within the peak region [cf.\ Eq.~\eqref{eq:relaxMode}].
Moreover, we used the diffusion- and reaction-limited rates $\sigma_\mathrm{D,R}$ of the peak-competition scenario (cf.\ Sec.~\ref{sec:peak-comp}).

The relaxation rate follows from the condition that the approximations in the peak region and in the plateau have to be matched.
Because the profile of the (long-wavelength and slowest) relaxation modes in the plateau only varies on the domain length and $b\ll \Lambda/2$, we use for matching that (sharp-peak approximation)
\begin{equation}
    \frac{\partial_x \delta\eta_\mathrm{plateau}|_{x=0}}{\delta\eta_\mathrm{plateau}|_{x=0}} \approx \frac{\partial_x \delta\eta_\mathrm{plateau}|_{x=b}}{\delta\eta_\mathrm{plateau}|_{x=b}}.
\end{equation}
Thus, setting Eqs.~\eqref{eq:relax-fraction-outer},~\eqref{eq:relax-fraction-inner} equal, one obtains
\begin{equation}\label{eq:relax-matching-cond}
    \tan(\zeta) \approx \zeta \left[ \frac{\Lambda\partial_{\etaStat} \rho_-}{-\partial_{\etaStat}M} + \frac{\sigma_\mathrm{D}^+}{\sigma_\mathrm{R}^+} \zeta^2\right]^{-1}.
\end{equation}
In the diffusion-limited regime ${\sigma_\mathrm{D}^+ \ll \sigma_\mathrm{R}^+}$ [cf.\ Secs.~\ref{sec:peak-coal} and \ref{sec:peak-comp}], the (first) intersection between the left-hand and right-hand sides will occur at ${\zeta \sim 1}$, which implies that the (smallest) diffusion-limited relaxation rate is
\begin{equation}
    \sigma_\mathrm{relax}^\mathrm{D} \sim \frac{4 T \chi_-\rho_-}{\Lambda^2} \partial_\rho \eta^*(\rho_-) = \sigma_\mathrm{D}^+ \frac{\partial_{\etaStat} M}{\Lambda \partial_{\etaStat} \rho_-}.
\end{equation}
Similarly, in the reaction-limited regime ${\sigma_\mathrm{R}^+ \ll \sigma_\mathrm{D}^+}$, the (first) intersection will occur for ${\zeta^2 \sim \sigma_\mathrm{R}^+/\sigma_\mathrm{D}^+ \ll 1}$.
Expanding Eq.~\eqref{eq:relax-matching-cond} to first order in $\sigma_\mathrm{R}^+/\sigma_\mathrm{D}^+$, one obtains that the intersection lies at 
\begin{equation}
    \sigma_\mathrm{relax}^\mathrm{R} 
    \approx 
    \sigma_\mathrm{R}^+ \frac{\partial_{\etaStat} M}{\Lambda \partial_{\etaStat} \rho_-}
\end{equation}
in the assumed limit ${|\partial_{\etaStat} M| \gg \Lambda \partial_{\etaStat} \rho_-}$ which is the limit in which both the diffusion- and the reaction-limited relaxation of the elementary peak pattern are fast compared to peak competition.

%

\end{document}